\documentclass[a4paper,fleqn]{cas-dc}

\usepackage[numbers]{natbib}
\usepackage{diagbox}
\usepackage{multirow}
\usepackage{ulem}
\usepackage{graphicx}
\usepackage{subfigure}
\usepackage{epsfig}

\def\tsc#1{\csdef{#1}{\textsc{\lowercase{#1}}\xspace}}
\tsc{WGM}
\tsc{QE}
\tsc{EP}
\tsc{PMS}
\tsc{BEC}
\tsc{DE}

\begin{document}

\begin{titlepage} 

	\centering 
	
	\scshape 
	
	\vspace*{\baselineskip} 
	
	
	\rule{\textwidth}{1.6pt}\vspace*{-\baselineskip}\vspace*{2pt} 
	\rule{\textwidth}{0.4pt} 
	
	\vspace{0.75\baselineskip} 
	
	{\LARGE Evaluatology: The Science and Engineering of Evaluation} 
	
	\vspace{0.75\baselineskip} 
	
	\rule{\textwidth}{0.4pt}\vspace*{-\baselineskip}\vspace{3.2pt} 
	\rule{\textwidth}{1.6pt} 
	
	\vspace{2\baselineskip} 
	
	
	
	\vspace*{3\baselineskip} 
	
	
	 \begin{center}Edited By\end{center}
	
	\vspace{0.5\baselineskip} 
	
	{ \begin{center} Jianfeng Zhan \\ Lei Wang \\ Wanling Gao \\ Hongxiao Li \\ Chenxi Wang \\ Yunyou Huang \\ Yatao Li \\ Zhengxin Yang \\ Guoxin Kang \\ Chunjie Luo \\ Hainan Ye \\ Shaopeng Dai \\ Zhifei Zhang \end{center}}
	
	
	\vspace{0.5\baselineskip} 

	\vfill 
	
	
	\epsfig{file=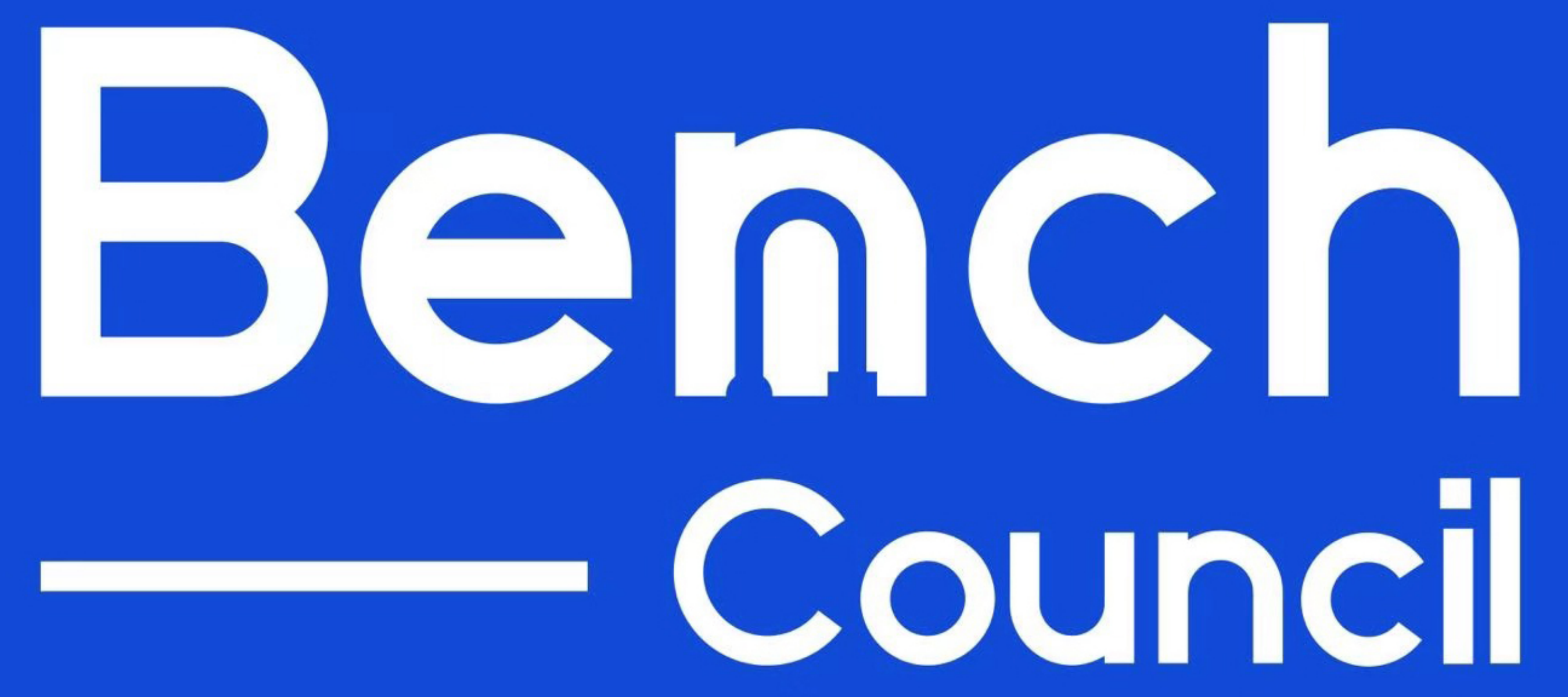,height=2cm}
	\textit{\\BenchCouncil: International Open Benchmark Council\\http://www.benchcouncil.org} 
	\vspace{5\baselineskip} 

	Technical Report No. BenchCouncil-Evaluatology-2024 
	
	{\large March 19, 2024} 

\end{titlepage}

\let\WriteBookmarks\relax
\def\floatpagepagefraction{1}
\def\textpagefraction{.001}
\shorttitle{Evaluatology: The Science and Engineering of Evaluation.}
\shortauthors{Prof. Dr. Jianfeng Zhan et al.}

\title [mode = title]{Evaluatology: The Science and Engineering of Evaluation}




\author[1,2,3]{Jianfeng Zhan}
\credit{Contribution to the proposal for the science and engineering of evaluation, including the universal evaluation concepts and terminologies, the universal evaluation theory and methodology, the universal benchmark concepts, principles, theories, and methodologies. Contribution to the mathematical formulation of fundamental evaluation issues. Contributions to the presentation of most texts, except figures and tables. Contributions to the other work unless stated explicitly by someone else}
\cormark[1]
\fnmark[1]
\ead{jianfengzhan.benchcouncil@gmail.com}
\ead[url]{www.zhanjianfeng.org}
\author[2,1,3]{Lei Wang}
\credit{Contributions to the whole-session discussion. The contribution to the axiom of the relative true evaluation results. The contribution to the CPU benchmark experiments in 5.2. The contributions to the summary of the related work for finance and HPC benchmarks in 5.1. Contribution to the mathematical formulation of fundamental evaluation issues}
\author[2,1,3]{Wanling Gao}
\credit{Contributions to the presentations of all figures. Contributions to the whole-session discussion. The discussion of the benchmark inspired the first author to think about what is essential of the benchmark}
\author[2,3]{Hongxiao Li}
\credit{Contributions to the whole-session discussion, mathematics notations, mathematical formulations of three fundamental issues in Evaluatology, and mathematical formulations of EEC and LEEC}
\author[2,3]{Chenxi Wang}
\credit{Contributions to the whole-session discussion. The contribution to the design, implementation, analysis, and presentation of experiments in 5.2}
\author[4,1]{Yunyou Huang}
\credit{Contribution to the discussion and the summary of the related work of drug evaluations in Section 5.1}
\author[6]{Yatao Li}
\credit{Contribution to the discussion and mathematical formulations of evaluation traceability in Section 3.7.3}
\author[2,3]{Zhengxin Yang}
\credit{Contribution to the discussion and presentation of metrology. Contribution to the discussion and real-world evaluation of safety-critical systems}
\author[2,1,3]{Guoxin Kang}
\credit{Contribution to discussions and the related work about business benchmarking in 5.1}
\author[2,1,3]{Chunjie Luo}
\credit{Contribution to the discussions}
\author[1,2,3]{Hainan Ye}
\credit{Contribution to the discussions}
\author[3]{Shaopeng Dai}
\credit{Contribution to the presentations of some figures}
\author[5,1]{Zhifei Zhang}
\credit{Contribution to the discussion and the summary of related work of drug evaluations}


\address[1]{The International Open Benchmark Council}
\address[2]{ICT, Chinese Academy of Sciences, Beijing, China}
\address[3]{University of Chinese Academy of Sciences, Beijing, China}
\address[4]{Guangxi Normal University, Guilin, Guangxi, China }
\address[5]{Capital Medical University, Beijing, China}
\address[6]{Microsoft Research Asia}

\begin{abstract}
Evaluation is a crucial aspect of human existence and plays a vital role in various fields. However, it is often approached in an empirical and ad-hoc manner, lacking consensus on universal concepts, terminologies, theories, and methodologies. This lack of agreement has significant repercussions. This article aims to formally introduce the discipline of evaluatology, which encompasses the science and engineering of evaluation. We propose a universal framework for evaluation, encompassing concepts, terminologies, theories, and methodologies that can be applied across various disciplines. 

Our research reveals that the essence of evaluation lies in conducting experiments that intentionally apply a well-defined evaluation condition to diverse subjects and infer the impact of different subjects by measuring and/or testing. Derived from the essence of evaluation, we propose five axioms focusing on key aspects of evaluation outcomes as the foundational evaluation theory. These axioms serve as the bedrock upon which we build universal evaluation theories and methodologies. When evaluating a single subject, it is crucial to create evaluation conditions with different levels of equivalency. By applying these conditions to diverse subjects, we can establish reference evaluation models. These models allow us to alter a single independent variable at a time while keeping all other variables as controls. When evaluating complex scenarios, the key lies in establishing a series of evaluation models that maintain transitivity. Building upon the science of evaluation, we propose a formal definition of a benchmark as a simplified and sampled evaluation condition that guarantees different levels of equivalency. This concept serves as the cornerstone for a universal benchmark-based engineering approach to evaluation across various disciplines, which we refer to as benchmarkology.

\end{abstract}

\maketitle

\section{Introduction}~\label{Section_Introduction}

\begin{figure}
\centering
\subfigure[The essence of evaluation.]{
\begin{minipage}[b]{0.42\textwidth}
\includegraphics[scale=.49]{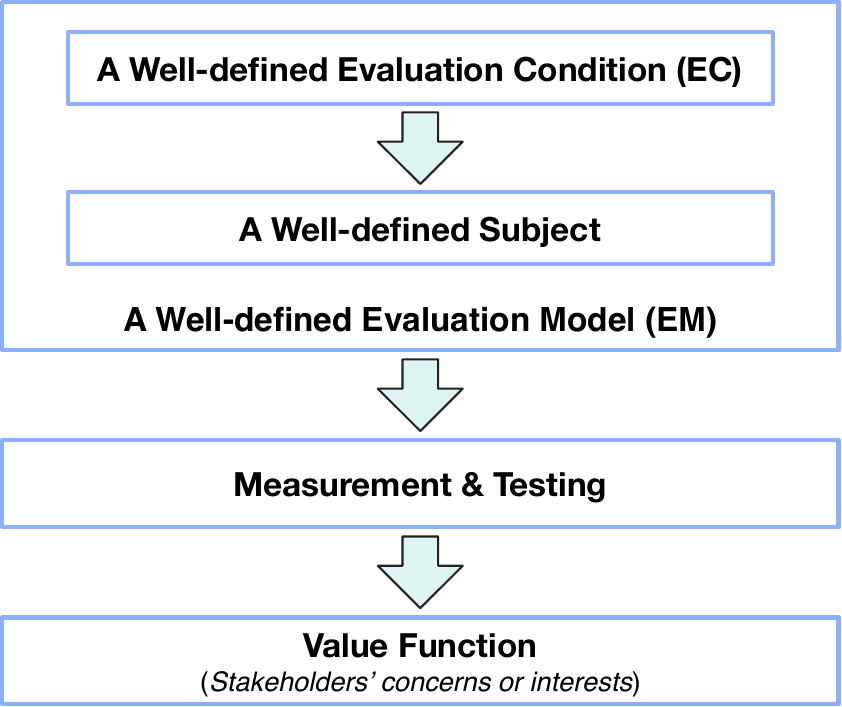} 
\end{minipage}
}
\subfigure[The basic methodology of evaluating a single subject.]{
\begin{minipage}[b]{0.42\textwidth}
\includegraphics[scale=.49]{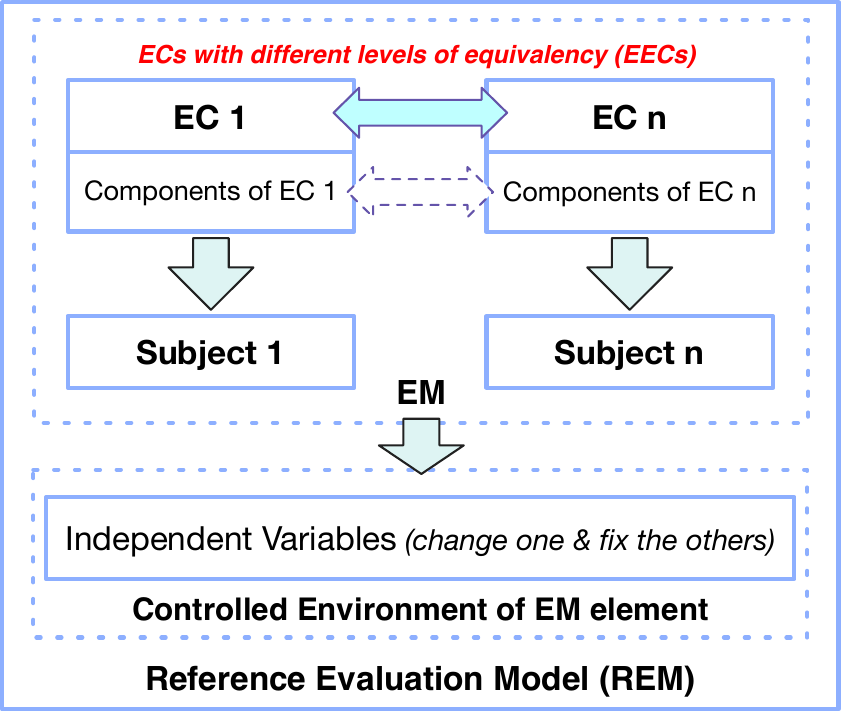}
\end{minipage}
}
\subfigure[The evaluation methodology addressing complexities that arise in more intricate scenarios.]{
\begin{minipage}[b]{0.42\textwidth}
\includegraphics[scale=.49]{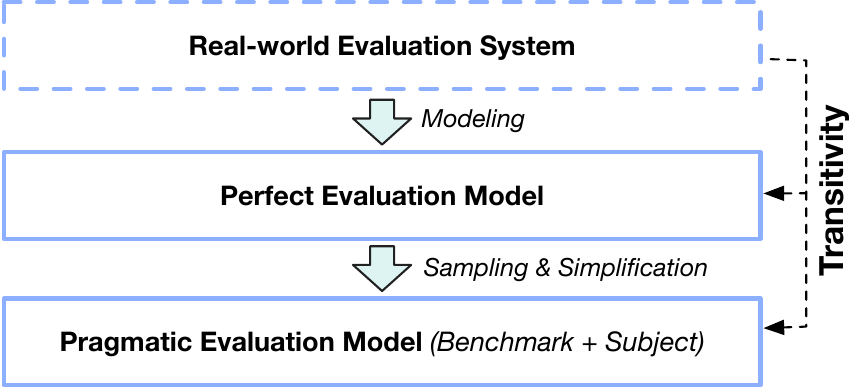}
\end{minipage}
}
\caption{The universal concepts, theories, and methodologies in evaluatology.}	\label{intr-evaluatology}
\end{figure}

Evaluation, a fundamental and significant undertaking in human existence, possesses a multifaceted nature. It spans a wide spectrum of domains, encompassing the assessment of computer performance, the evaluation of societal interventions to determine their efficacy~\cite{rossi2018evaluation}, the ranking of educational institutions, and even the appraisal of political leaders through electoral processes. As a result, evaluation assumes a pivotal role that permeates every discipline. 
Nevertheless, it is pertinent to recognize that evaluation practices often adopt ad-hoc and empirical approaches, displaying inherent variations among various disciplines.


 Collectively, evaluations within diverse disciplines lack universal concepts, terminologies, theories, and methodologies. In the disciplines of computer science, social sciences, and psychology, the communities develop different methodologies that design experiments to deliberately impose treatments on individuals or systems under scrutiny, which we refer to as the \textit{subject}, to measure and analyze their responses~\cite{statics_book}. In the field of computer science, a \textit{benchmark} is utilized as a tool and  methodology~\cite{john2018performance,hennessy2011computer, browne1975analysis, knudson1985performance} to evaluate the effectiveness and efficiency of system designs and implementations. 
 In the realm of social sciences, evaluation assumes the application of social research methodologies to systematically investigate the effectiveness and efficiency of intervention programs aimed at enhancing societal conditions, as defined by Rossi et al.~\cite{rossi2018evaluation}. Within the psychology domain, social and personality psychologists often employ \textit{scales} such as psychological inventories, tests, or questionnaires~\cite{furr2011scale} to quantify psychometric variables~\cite{furr2011scale}.  
 
 Conversely, evaluations in the business, finance, and education domains take different observational study methodologies~\cite{statics_book}.  The field of business embraces the concept of \textit{benchmarking} as a means to identify exemplary practices that can propel companies towards superior performance~\cite{camp1989benchmarking}. 
In the realms of finance and education, evaluation often utilizes a tool named \textit{index}. Widely employed to gauge the overall performance of the system, the index is derived through the meticulous calculation of the weighted average, utilizing a select group of representative individuals~\cite{fisher1966some}.

Discussions concerning the true essence of evaluation are seldom found, which often results in confusion with measurement and testing and lacks clear differentiation. A critical consequence of this absence is the lack of previous endeavors to establish universally applicable foundational evaluation principles and methodologies that cut across diverse disciplines, ultimately giving rise to significant ramifications.


Even within computer sciences and engineering, it is not uncommon for evaluators to generate greatly divergent evaluation outcomes for the same subject. These discrepancies can range from significant variations to the extent of yielding contradictory qualitative conclusions. An example of this phenomenon can be observed when using multiple widely recognized CPU benchmark suites to assess the performance of the same processor. This often leads to greatly divergent evaluation outcomes that are \textit{incomparable} across different benchmarks.
 Such circumstances give rise to valid concerns surrounding the reliability, effectiveness, and efficiency of these approaches when appraising the subject that is critical to safety, missions, and businesses. Further details on this issue can be found in Section~\ref{SubSection_Motivation}.




To the best of our knowledge, this article, for the first time, formally introduces the discipline of evaluatology, encompassing the science and engineering of evaluation. We present an all-encompassing concept, terminology, theory, and methodology framework for evaluation that can be universally applied across diverse disciplines.



We highlight that the essence of evaluation lies in conducting deliberate experiments where a well-defined Evaluation Condition (EC) is applied to a subject to establish a well-defined Evaluation Model (EM). Subsequently, we can infer the impacts of the subjects by measuring and/or testing an EM. 
Derived from the core essence of evaluation, we present five axioms as the foundational principles of evaluation theory.  The five axioms focus on key aspects of evaluation outcomes, including true quantity (The first and second axioms), traceability of discrepancy (the third axiom), comparability (the fourth axiom), and realistic estimate (the fifth axiom).

        
      
      
   


Based on the five evaluation axioms, we present the universal evaluation theories and methodologies from two distinct dimensions: evaluating a single subject and complex scenarios.

A well-defined EC serves as a prerequisite for meaningful comparisons and analyses of the subjects. We propose a universal hierarchical definition of an EC and identify five primary components of an EC from the top to the bottom. 

In the process of evaluating subjects, it is of utmost importance to prioritize the use of the \uline{e}quivalent ECs (EECs) across diverse subjects. This means that in order to establish two EECs, it is crucial to ensure that the corresponding components within the same layer of the two ECs are equivalent. By maintaining equivalency at each layer, we can ensure fair and unbiased evaluations, enabling meaningful comparisons and assessments between different subjects.

In certain cases, achieving complete equivalence between two ECs at all levels can be a challenging or even unattainable task. In such cases, we propose a minimum requirement of ensuring uniformity in the most essential components of the two ECs, which we refer to as the least equivalent evaluation condition (LEEC). To establish the LEEC, we identify the most governing component within an EC that must exhibit equivalency. This component, known as the evaluation standard, plays a crucial role in defining the LEEC. 

We apply ECs with different levels of equivalency to diverse subjects to constitute EMs. An EM element refers to a specific point within the EM state space, and each EM element may have many independent variables. To eliminate confounding, we propose a new concept named a reference evaluation model (REM). An REM mandates that each element of an EM change only one independent variable at a time while keeping the other independent variables as controls. Subsequently, we utilize the measurement and/or testing to gauge the functioning of the REM. Finally, from the amassed measurement and testing data of the evaluation systems, we then deduce the cause-effect impacts of the different subjects.

Addressing the complexities that arise in more intricate scenarios, we reveal that the key to effective and efficient evaluations in various complex scenarios lies in the establishment of a series of EMs that maintain transitivity. 

 In real-world settings, we refer to the entire population of real-world systems that are used to evaluate specific subjects as the \textit{real-world evaluation system (ES)}. Assuming no safety concerns are present, the real-world ES serves as a prime candidate for creating an optimal evaluation environment, enabling the assessment of diverse subjects. However, there are several significant obstacles to consider, i.e., the presence of numerous confounding, the challenges of establishing an REM, high evaluation costs resulting from the huge state spaces, multiple irrelevant concurrent problems or tasks taking place, and the tendency to bias specific clusters of EC state space.

We posit the existence of a \textit{perfect EM} that replicates the real-world ES with utmost fidelity. A perfect EM eliminates irrelevant problems or tasks, supports free and artificial settings of different evaluation configurations, facilitates the establishment of REMs, and has the capability to thoroughly explore and comprehend the entire spectrum of possibilities of an EC.
However, the perfect EM possesses huge state space, entails a vast number of independent variables, and hence results in huge evaluation costs. To address this challenge, it is crucial to propose a pragmatic EM that simplifies the perfect EM in two ways: reducing the number of independent variables that have negligible effect and sampling from the extensive state space.  A pragmatic EM provides a realistic estimate of the parameters of the real-world ES.

We put forth four fundamental issues in the evaluations and formally formulate the problems mathematically: ensure the transitivity of EMs; perform a cost-efficient evaluation with controlled discrepancies; ensure the evaluation traceability; connect and correlate evaluation standards across diverse disciplines.



Building upon the science of evaluation, we formally define a benchmark as a simplified and sampled EC, specifically a pragmatic EC, that ensures different levels of equivalency.  Based on this concept, we propose a benchmark-based universal engineering of evaluation across different disciplines, which we aptly term "benchmarkology."  Figure~\ref{intr-evaluatology} presents the universal concepts, theories, and methodologies in Evaluatology.  

The article is structured as follows: Section~\ref{background_work} elucidates the background.
Section~\ref{evaluationonology} introduces a comprehensive theoretical and methodological framework for evaluatology. Section~\ref{benchmarkology} outlines the principles and methodologies of benchmarkology. 
Section~\ref{SubSection_Motivation} reviews the state-of-the-art and state-of-the-practice evaluations and benchmarks and expounds upon the imperative to cultivate the science and engineering of evaluation. 
Ultimately, Section~\ref{conclusion} manifests the overarching conclusion.





\section{Background}\label{background_work}

This section provides an overview of the background.

\subsection{Basic concepts}~\label{basic_concept_definition}

This subsection presents several concepts, like individual, systems, populations, samples, variables,   models,  confounding, control, and treatment based on several undergraduate textbooks~\cite{statics_book,physics_book,calculus_stewart2018single}. 

 
\textit{An individual} can be defined as the object that is described by a given set of data.  A \textit{system} is an interacting or interdependent group of individuals, whether of the same or different kinds,  forming a unified whole~\cite{system_definition,backlund2000definition}. A system could be a recursive structure. That is to say, a high-level system could consist of an interacting or interdependent group of low-level systems, whether of the same or different kinds,  forming a unified whole. 

A \textit{population} is the entire group of individuals or systems we wish to study and understand, while a \textit{sample} represents a smaller subset of individuals or systems from the population~\cite{statics_book}.  A \textit{variable} or \textit{quantity} is any \textit{property} of an individual or system. 
A \textit{parameter} is a number that describes some property of the population, while a \textit{statistic} is a number that describes some property of a sample.
\textit{Inference} is the process of drawing conclusions about a parameter of a population on the basis of the statistic of sample data~\cite{statics_book}.

According to ~\cite{calculus_stewart2018single}, a function, denoted as f, is a rule that assigns a unique element, referred to as $f(x)$, from a set $R$ to each element in a set $D$. In this context, the domain, denoted as $D$, refers to the set of all possible values for which the function is defined. On the other hand, the range of the function, denoted as $f(x)$, consists of all the possible values that $f(x)$ can take as $x$ varies within the domain. The \textit{independent variable} is represented by a symbol that encompasses any arbitrary number within the domain of the function.   A \textit{dependent variable}, represented by a symbol, is used to denote a number within the range of the function.

A \textit{model} is a simplified version of a system
that would be too complicated to analyze in full detail~\cite{physics_book}. A model could be a physical model or a \textit{mathematical model}. A mathematical model is a mathematical description, typically through \textit{functions} or equations, of a system, the purpose
of which is to understand the system and to make predictions about its behavior~\cite{calculus_stewart2018single}. Throughout the remainder of this article, we will use the terms "system" and "model" interchangeably unless explicitly stated otherwise.

A \textit{treatment} refers to a specific condition or intervention that is applied to the individuals or systems under study. 
An \textit{observational study} observes individuals or systems and measures variables of interest without any attempt to influence their responses, while an \textit{experiment} is designed to deliberately impose treatment on individuals or systems to measure and analyze their responses~\cite{statics_book}. In experiments with multiple independent variables, a treatment is a combination of specific values assigned to these variables. 

In the realm of understanding cause and effect, it is crucial to rely on experiments rather than observational studies. Even if an observational study is based on a random sample, it still falls short in effectively measuring the impact of changes in one variable on another variable~\cite{statics_book}. Experiments, on the other hand, provide us with compelling and conclusive data, making them the only source that truly convinces us of cause-and-effect relationships.


\textit{Confounding} arises when two independent variables are associated in a manner that makes it challenging to differentiate their specific effects on a dependent variable. In other words, the influence of these independent variables becomes entangled, making it difficult to attribute specific impacts to each one. In such cases, the independent variable responsible for this confounding effect is referred to as a \textit{confounding variable}.  \textit{Control} means keeping other independent variables that might affect the response the same~\cite{statics_book}, and the main purpose of a  \textit{control group} is to provide a baseline for comparing
the effects of the other treatments.

\subsection{Metrology}

Metrology is the science of measurement and its applications~\cite{bipm2012international}. In this section, we present a simplified yet systematic framework for understanding metrology concepts based on the works of~\cite{bipm2012international, kacker2021quantity}. To maintain conciseness, we focus only on the essential metrology concepts.

\begin{figure}
	\centering
		\includegraphics[scale=.55]{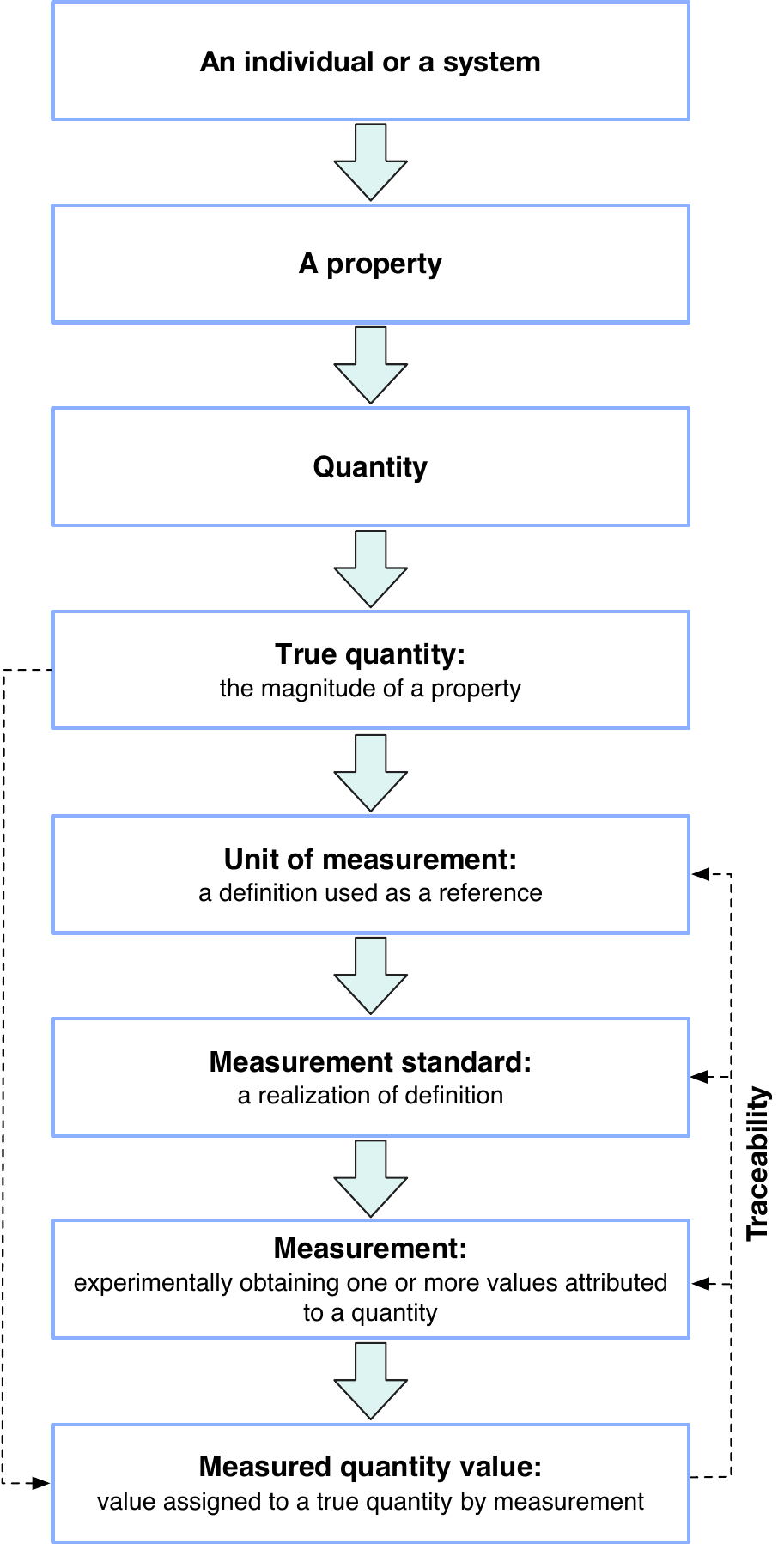}
	\caption{A simplified yet systematic conceptual framework for metrology~\cite{bipm2012international, kacker2021quantity}. }
	\label{metro-fram}
\end{figure}

The essence of metrology lies in quantities and their corresponding measurements. A \textit{quantity} is a property whose instances can be compared by ratio or only by order~\cite{bipm2012international}. 
Furthermore, Psychologist Stanley Smith Stevens developed a well-known measurement classification with four levels based on empirical operations, mathematical group structure, and permissible statistics (invariant): nominal, ordinal (based on order~\cite{bipm2012international}), interval, and ratio~\cite{stevens1946theory}~\footnote{In the original article, Stanley Smith Stevens used the term "levels or scales of measurement." We have only used "levels" to avoid confusion with the specific meaning of "scales" in psychology}.

A nominal level is the most basic form of measurement, where numbers are used as labels or type numbers to establish an equality relation. An ordinal level, on the other hand, involves ranking the items in a particular order.  An interval level exhibits an equality of interval relation, where (1) the choice of a zero point is a matter of convention or convenience, (2) there is rank ordering, and (3) the scale remains invariant when a constant is added to all values, preserving the differences between them. A ratio level allows for all four types of relations: equality, rank-ordering, equality of intervals, and equality of ratios.

The international system of metrology encompasses seven fundamental quantities: time, length, mass, electric current, thermodynamic temperature, amount of substance, and luminous intensity~\cite{bipm2012international}. 

Consistent with the definition of a quantity, the true value of a quantity represents the magnitude of a property or characteristic of an individual or system, e.g., a phenomenon, body, or substance that is independent of any observer. For example, it can be a specific circle's radius or a particular particle's kinetic energy within a given system ~\cite{kacker2021quantity,bipm2012international}.  For measurement, the true quantity value is an unknown measurement target~\cite{bipm2012international}.


In the field of measurement, the \textit{unit of measurement}~\cite{kacker2021quantity} plays a crucial role. It is a real scalar magnitude that is defined and adopted by convention. Its purpose is to allow for the comparison of quantities of the same kind.

\textit{Measurement standard}~\cite{kacker2021quantity} is a realization of the definition of quality. It is characterized by a stated metric value and an associated measurement uncertainty.

To establish a measurement standard, it is important to use a \textit{measurement methodology} that is both repeatable (performed by the same team) and reproducible (performed by different teams). This ensures consistency and reliability in the reference for measurements. Such measurements can be conducted using \textit{measuring instruments} or \textit{measuring systems}~\cite{bipm2012international}, providing a reliable foundation for further analysis and comparison.

\textit{Measurement} is experimentally obtaining one or more values attributed to a quantity and other relevant information~\cite{bipm2012international}. Another widespread definition of measurement in the social sciences is "the assignment of numerals to objects or events according to some rule."~\cite{stevens1946theory}, dating back to 1946.  
Quantity values obtained by the measurement are measured (quantity) values, representing the measurement results~\cite{bipm2012international}.





The hierarchy of measurement standards follows a progression from lower to upper levels, with increasing accuracy and cost. This progression starts from national measurement standards and extends to international standards. As a property of a measurement result, \textit{measurement traceability}~\cite{bipm2012international} establishes a connection between the result and a reference (measurement standards, measuring instruments, and measuring systems). This connection is established through a documented, unbroken chain of calibrations, with each calibration contributing to the measurement uncertainty. To ensure accuracy, each level of measurement standards in the hierarchy should be calibrated using a higher standard with greater precision.




 
\subsection{Testing}~\label{testing}


A \textit{test oracle} is a method used to verify whether an individual or system being tested has performed correctly during a specific execution~\cite{baresi2001test}. Test oracles can include, but are not limited to, specifications, contracts, reference products, previous versions of the same product, and relevant performance or quality of service criteria~\cite{choudhary2011software}.

\begin{figure}
	\centering
		\includegraphics[scale=.45]{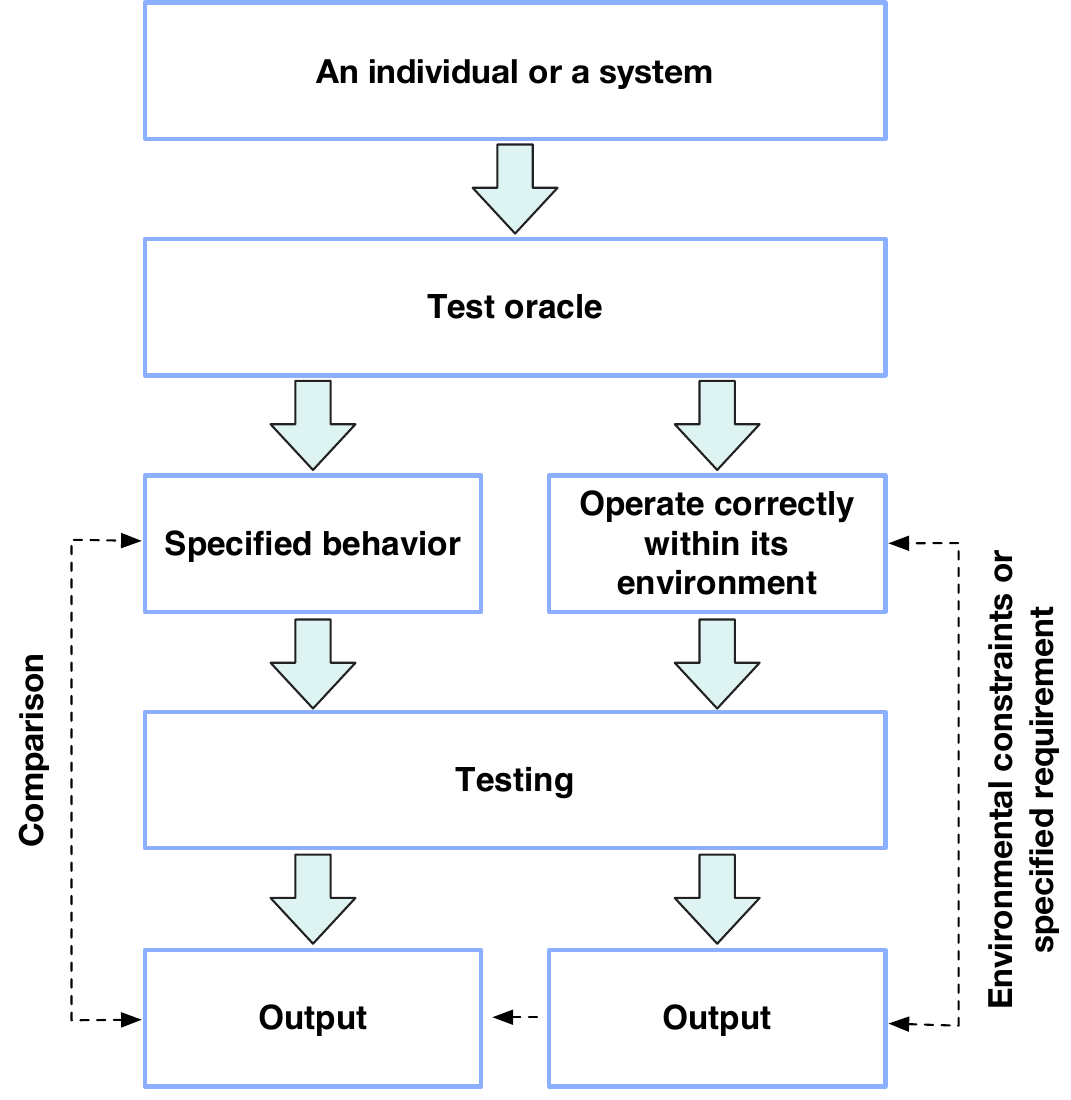}
	\caption{A simplified yet systematic conceptual framework for testing~\cite{baresi2001test,whittaker2000software}. }
	\label{testing-fram}
\end{figure}

\textit{Testing} is the process of executing an individual or system to determine whether it (1) conforms to the specified behavior defined by the test oracles~\cite{whittaker2000software} (the first category) and/or (2) operates correctly within its intended environment as defined by the test oracles (the second category).


In the first category of testing, the test oracle compares the actual output with the specified output to identify incorrect behavior, which is considered a \textit{failure}~\cite{ whittaker2000software}. Another type of failure is often encountered in the second category of testing, where an individual or system fails to meet environmental constraints or falls outside the specified requirements. Examples of such failures include running out of memory, slow execution, and incompatibility with operating systems ~\cite{baresi2001test}. It is important to note that these two types of failures are not isolated incidents. Failures in the second category can lead to failures in the first category, such as running out of memory, which results in incorrect program execution.

\section{The science of evaluation}~\label{evaluationonology}

Within this section and the forthcoming one, we first propose the universal evaluation concepts and terminologies. Then, we elucidate the theory and methodology governing the realm of evaluation, collectively referred to as evaluatology. 

We present the science of evaluation from two distinct perspectives: evaluating a single subject and evaluating complex scenarios.






We postpone the exposition of the principles and methodology pertaining to benchmarkology, the benchmark-based engineering of evaluation, to Section~\ref{benchmarkology}.








\subsection{Basic evaluation concepts and terminology}~\label{basic_concepts}

In this subsection, we will examine an illustrative case study in evaluation and meticulously analyze the fundamental \textit{components} inherent in an evaluation. For the sake of convenience, henceforth, within this article, each case study shall be assigned a unique case ID for differentiation purposes.


In the first case, which we'll call Case One, an organization is in the process of acquiring a computer. To make an informed decision, the organization decides to evaluate various computer options by executing its applications on each one. During this evaluation, the organization will collect extensive data on performance and energy efficiency.

Based on this data, the organization will then formulate an explicit or implicit function to express its preferences for different computers. This function will serve as a way to quantify and articulate the organization's priorities and requirements in terms of performance and energy efficiency. By doing so, the organization can make a well-informed decision and choose the computer that best aligns with its needs and preferences.


In any evaluation process, we refer to an individual or a system under scrutiny as a \textit{subject}. In the context of Case One, the computers that are being considered for evaluation are the subjects.

Another important component of an evaluation is the presence of \textit{stakeholders}. A stakeholder is defined as an entity that holds a stake of responsibility or interest in the subject matter.  In this case, the stakeholders involved in the procurement of computers include the organization seeking to acquire the computers, the specific users, the designers responsible for creating the computer specifications, and the producers who manufacture the computers.

Each subject, in this case, the computer, has its own set of stakeholders who will render judgments based on the data collected from measurements. Metrology provides the foundational principles, methodology, and instruments for measurements within this process. It ensures that the measurements are accurate, reliable, and consistent, enabling the stakeholders to make informed decisions based on the gathered data.



In Case One, the prospect of measuring the mere attributes of a subject, such as its weight and power consumption, possesses a certain degree of utility. Nevertheless, such measurements fall considerably short of meeting stakeholders' evaluation requirements. The stakeholder seeks comprehension of the subject's effectiveness and efficiency when applying a specific condition or intervention to the subject. In this case, it is to execute the stakeholder's primary or forthcoming applications, which we informally label as an \textit{evaluation condition (EC)}.   Other disciplines commonly use the term "treatment" to refer to the same concept. The EC represents the third critical component of an evaluation, which shall be formally expounded upon subsequently.



Within this framework, an important question arises: How can organizations establish a framework to determine the preferences of distinct subjects when they exhibit varying levels of performance across different applications?


In the current state-of-the-practice, a more intuitive approach entails executing applications on computers sequentially. Subsequently, we proceed to measure the computers' performance when operating distinct applications individually. Following each execution, data is collected encompassing factors such as the duration of each application's execution and its corresponding energy consumption.

It is imperative to establish a function that can map the compiled measurement data to one or several composite evaluation metrics capable of capturing the stakeholders' concerns and interests. In the rest of this article, we refer to this function as \textit{a value function}. Once the evaluation outcomes have been obtained, it becomes feasible to define a \textit{reference subject} and its \textit{reference evaluation outcome} against which the evaluation results of alternative subjects can be compared.

\subsection{The essence of evaluation}~\label{essence_of_an_evaluation}

From the aforementioned analysis in Section~\ref{basic_concepts}, it is evident that the challenge in evaluation arises from the inherent fact that each subject cannot be evaluated in isolation. Evaluating a subject in isolation falls short of meeting the expectations of stakeholders. Therefore, it is crucial to apply a well-defined EC directly to the subjects. An EC consists of multiple components, and a well-defined EC is one where all its composing elements are clearly defined. In essence, evaluation can be viewed as an intentional experiment that deliberately imposes a specific EC on the subject itself.


Based on the definitions provided in Section~\ref{basic_concept_definition}, when a subject is equipped with an EC, it forms an \textit{evaluation system} (ES) or an \textit{Evaluation Model} (EM).  An evaluation model (EM) is a simplified version of an ES that would be too complicated to analyze in full detail~\cite{physics_book}. 

Based on the analysis presented earlier, it becomes clear that \textit{the core essence of evaluation lies in conducting deliberate experiments where EECs are applied to a diverse range of subjects, resulting in the establishment of equivalent EMs. Subsequently, we can effectively evaluate the subjects by measuring the equivalent EMs.}

Therefore, we formally define evaluation as \textit{an experiment that applies EECs to diverse subjects and establishes equivalent EMs, enabling the measurement of these equivalent EMs, the inference of the subjects' impact, and the subsequent judgment of them}.






\subsection{Five evaluation axioms}\label{evaluation_axiom} 

In this section, we present five evaluation axioms that are derived from the core essence of evaluation, serving as the foundational principles of evaluation theory. They focus on key aspects of evaluation outcomes, including true quantity (The first and second axioms), traceability of discrepancy (the third axiom), comparability (the fourth axiom), and realistic estimate (the fifth axiom).



\textbf{The First Axiom of Evaluation: The Axiom of the Essence of  Composite Evaluation Metrics.} This axiom declares that the essence of the composite evaluation metric either carries inherent physical significance or is solely dictated by the value function. 

In nature, \textit{a composite evaluation metric refers to a combined quantity that is constructed using base quantities and other quantities that possess physical significance.}
If a composite evaluation metric does not carry inherent physical significance, the value function serves as a mechanism that maps base quantities and other quantities carrying physical meaning into a composite evaluation metric. The widespread acceptance of the composite evaluation metric relies on it being embraced by the community of evaluators.  

\textbf{The Second Axiom of Evaluation: The Axiom of True Evaluation Outcomes.} This axiom declares that when a well-defined subject is applied to a well-defined EC, its evaluation outcomes, including its quantities and composite evaluation metrics, possess true values.

"The magnitude of a property of an individual phenomenon, body, or substance is associated with a true quantity"~\cite{kacker2021quantity,bipm2012international}. Additionally, each testing procedure yields a definitive outcome relative to its respective test oracle. Building upon this inference, it is reasonable to presume that when a well-defined subject is equipped with a well-defined EC, the quantities within the corresponding well-defined EM possess true values.
  
For a well-defined EM, each composite evaluation metric is derived from measurement and/or testing outcomes, utilizing a definite value function that translates the base quantities and other quantities into a composite evaluation metric. Consequently, the evaluation results are exclusively determined by the measurements and/or testing procedures employed.  It is reasonable to assume that for a well-defined EM, its composite evaluation metric possesses a true value.  

The first and second axioms are concerned with the true evaluation outcome. The first axiom provides the basis for defining value functions. With a well-defined value function, a well-defined EM possesses true quantities of evaluation outcomes.   



\textbf{The Third Axiom of Evaluation: The Axiom of Evaluation Traceability.} This axiom declares that for the same subject, the divergence in the evaluation outcomes can be attributed to disparities in ECs, thereby establishing evaluation traceability. This axiom focuses on the traceability of discrepancies in the evaluation outcomes.

For the same subject, this axiom is deemed rational as disparities in evaluation outcomes can be rationalized as the consequence of variations in the ECs. In the absence of this axiom, the differences observed in evaluation outcomes would be inexplicable, contradicting our scientific and engineering intuitions.

\textbf{The Fourth Axiom of Evaluation: The Axiom of Comparable Evaluation Outcomes.} This axiom declares when each well-defined subject is equipped with  EECs,  their evaluation outcomes are comparable. It goes without saying this axiom is related to the comparability of the evaluation outcomes. 

Only when each EC is well-defined, and two ECs achieve complete equivalence at all levels can we refer to them as EECs. When each well-defined subject is equipped with EECs, its evaluation outcomes possess true values. Additionally, when well-defined subjects are subjected to EECs, their evaluation outcomes accurately reflect the impacts of different subjects under the same conditions, making them comparable. 

\textbf{The Fifth Axiom of Evaluation: The Axiom of Consistent Evaluation Outcomes.} This axiom asserts that when a well-defined subject is evaluated using different samples from a population of ECs, their evaluation outcomes consistently converge toward the true evaluation outcomes of the population of ECs. This axiom provides a realistic estimate of the true evaluation outcomes. 

According to the Second Axiom of Evaluation, when a well-defined subject is equipped with a well-defined population of ECs, the resultant EM possesses the true evaluation outcomes.  When a sample is taken from a population of ECs, it serves as an approximation of the entire population. As a result, different samples yield consistent evaluation outcomes that gradually converge toward the evaluation outcomes of the entire population of ECs. This convergence is influenced by the sample's ability, which is determined by the chosen sampling policy, to represent the underlying characteristics of the population accurately.

\subsection{Basic evaluation  methodology}~\label{BEM}

This section outlines the fundamental methodology for evaluating a single subject. Drawing upon the discussion of the essence of evaluation in Section~\ref{essence_of_an_evaluation}, we propose a rigorous evaluation methodology to determine the impacts of the subjects as follows.

\begin{figure}
	\centering
		\includegraphics[scale=.6]{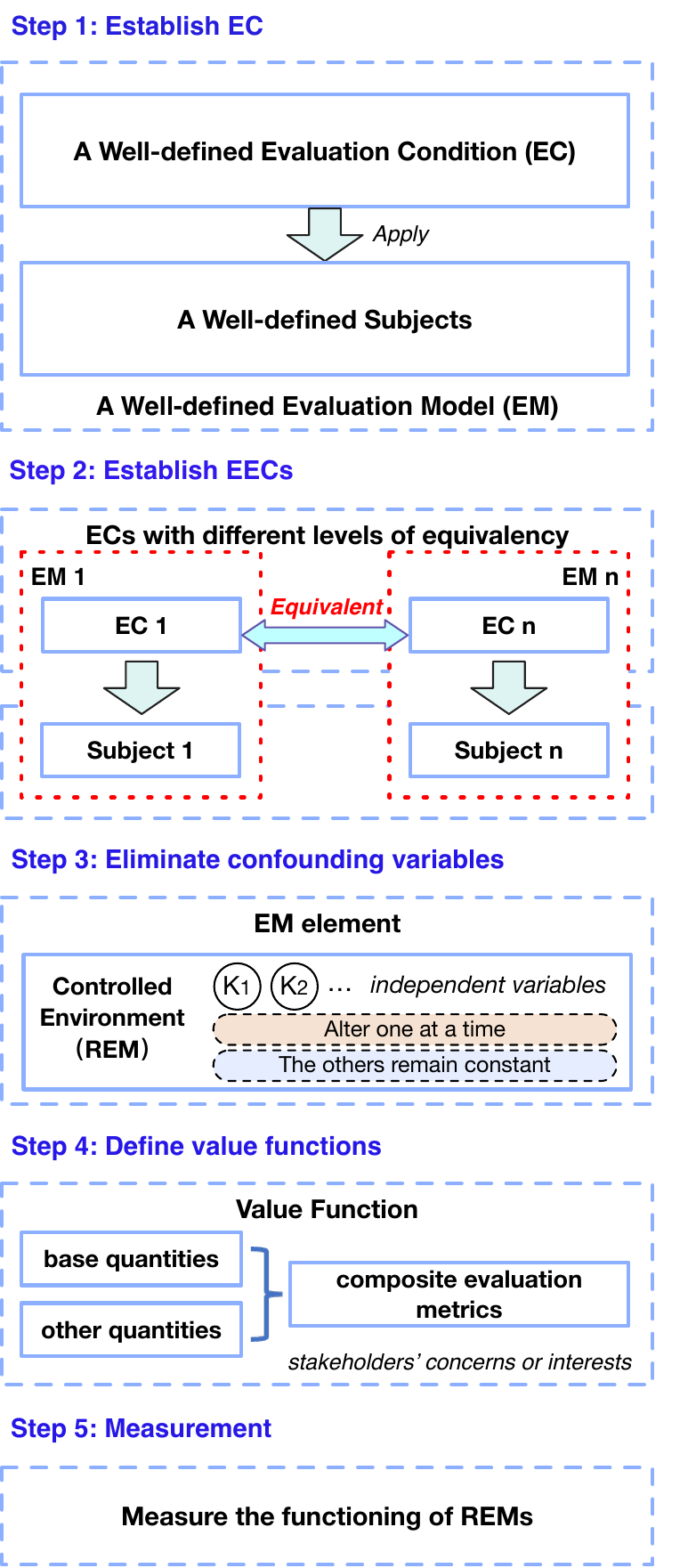}
	\caption{Basic evaluation methodology.}	\label{basic-evaluation-methodology}
\end{figure}

We create \uline{e}quivalent ECs (EECs) and apply EECs to diverse subjects to constitute equivalent EMs. An EM element refers to a specific point within the EM state space, and each EM element may have many independent variables. To eliminate confounding, we propose a new concept named a reference evaluation model (REM). An REM mandates that each element of an EM change only one independent variable at a time while keeping the other independent variables as controls. Subsequently, we utilize the measurement to gauge the functioning of the REM. Finally, from the amassed measurement data of the evaluation systems, we then deduce the cause-effect impacts of the different subjects. 

In this methodology, we emphasize five essential steps to ensure a comprehensive evaluation, as shown in Figure~\ref{basic-evaluation-methodology}. These steps are crucial in accurately determining the impacts of the subjects.

The first step is to establish a rigorous definition of an EC. According to The Second and Three Axioms of Evaluation, when a well-defined subject is applied to a well-defined EC, its evaluation outcomes possess true values; when the subject is applied to different ECs, any divergence in the evaluation outcomes can be attributed to disparities in these ECs. Therefore, the key focus in this phase is to present a well-defined EC clearly.

 The second step involves the establishment of EECs. As outlined in the Fourth Axiom of Evaluation, when EECs are applied to diverse subjects, their evaluation outcomes become comparable. Since the primary objective of evaluation is to compare different subjects, the establishment of EECs becomes an essential step in the process.

The third step involves the elimination of confounding variables. Given that each element of an EM consists of multiple independent variables, it becomes essential to establish an REM. An REM serves as a controlled evaluation environment where only one independent variable is altered at a time while the other independent variables remain constant as the control. This approach helps in isolating the effects of individual variables and ensures a more accurate evaluation of the subject's performance.

The fourth step is to define the value functions that map the base quantities and other quantities to composite evaluation metrics that represent the stakeholders' primary concerns or interests. According to the first axiom of evaluation, the essence of the composite evaluation metric either carries inherent physical significance or is solely dictated by the value function.  So, when defining a value function, it is crucial to make it become the consensus of the community. 

Finally, we utilize the measurement to gauge the functioning of the REM. From the amassed measurement data of the REM, we then deduce the cause-effect impacts of the different subjects.


\subsection{Basic evaluation  theory}~\label{BET}

This subsection presents the basic evaluation theory, including the hierarchical definition of an EC, universal concepts across different disciplines, the establishment of EECs, LLECs, evaluation standards, and the establishment of an REM.

\subsubsection{The hierarchical definition of an EC}\label{EC_definition}



In the preceding subsection, we deduced that there exists solely one feasible approach to evaluation: applying the EECs to diverse subjects and establishing an REM. Regrettably, even within a rudimentary evaluation setting such as Case One, an EC has multifarious components.



To address the above challenges, in this subsection, we propose a hierarchy definition of an EC. 
We start defining an EC from the problems or task spaces that these stakeholders face and need to address with the following two reasons. First,  the concerns and interests of the relevant stakeholders are at the core of the evaluation. These concerns and interests are best reflected through the problems or tasks they must face and resolve, which provide a reliable means to define an EC. Second, utilizing the same problem or task can ensure the comparability of evaluation outcomes.  


Taking Case One as an example, we observe that in Case One, an EC encompasses numerous constituents. Notably, we identify four primary components of an EC from the top to the bottom.
The first top component is a set of equivalent definitions of problems or tasks.  While the problem or task itself serves as the foundation for the evaluation process, it cannot solely serve as the evaluation itself because the problem or task is often abstract and requires further instantiation to determine its specific parameters.  The second component is the set of a collective of equivalent problem or task instances, each of which is instantiated from the element of the first component. Different from the first component, an equivalent problem or task instance is specific and could serve as the evaluation directly. 

After a problem or task instance is proposed, it is necessary to figure out a solution.  The third component consists of the algorithms or algorithm-like mechanisms, each of which provides the solution to a specific problem or task instance.  An algorithm-like mechanism refers to a process that operates in a manner similar to an algorithm. This term is proposed because, in numerous disciplines, such as social and biological sciences, it is not currently feasible to formulate mathematical algorithms explicitly. These domains often involve complex and nuanced phenomena that defy precise mathematical modeling.  

The fourth component encompasses the implementation of an algorithm or instantiation of an algorithm-like mechanism. Its implementation or instantiation involves understanding the algorithm or the algorithm-like mechanism and implementing it in a specific system. This process ensures that the algorithm or the algorithm-like mechanism can effectively and efficiently solve the intended problem instance or perform the desired task instance within the given context. 
  



In addition to the four components of an EC that we discussed above, other components can be involved in the other complex evaluation scenarios, which we will discuss later. 




\subsubsection{Universal concepts across different disciplines} \label{revisiting}

\begin{figure}
	\centering
		\includegraphics[scale=.5]{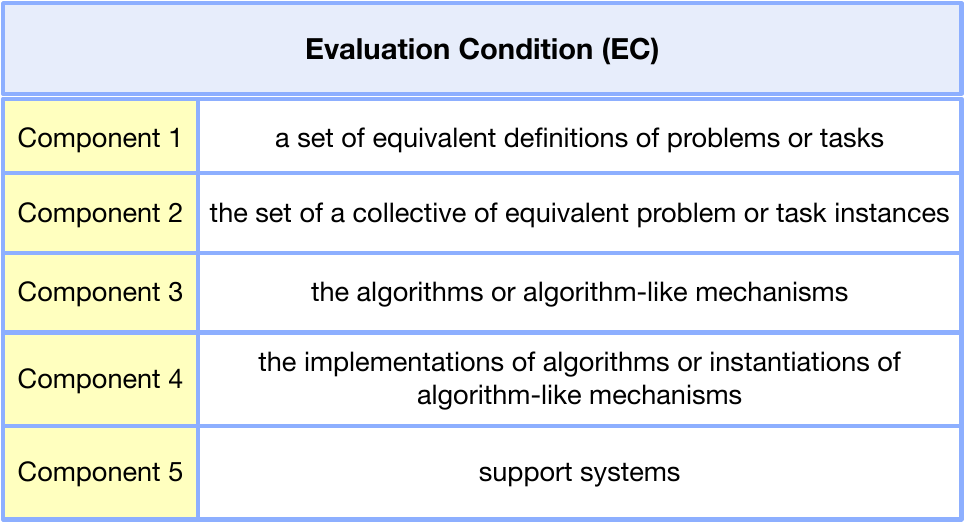}
	\caption{The Hierarchical Definition of an EC.}\label{ec-definition}
 \end{figure}

In Section~\ref{basic_concepts},  we conducted an examination
of the fundamental constituents of an evaluation, which
encompass "subjects" and "evaluation conditions." Additionally,
we put forth definitions for various fundamental concepts,
namely "subject," "stakeholders," and "value functions."
Through this analysis, we unveiled that the core nature of evaluation is to intentionally apply EECs to diverse subjects and establishan REM to infer the impact of the subjects for judgments.

However, given the multiplicity of evaluation scenarios,
two critical questions must be addressed: (1) Do these concepts
suffice for diverse scenarios? (2) Can we formulate a
comprehensive and universally applicable conceptual framework?
In this subsection, we delve deeper into various evaluation
cases across diverse disciplines, with the primary aim of enhancing
our comprehension of the evaluation process.

\textbf{Evaluating an AI algorithm}

In Case Two, the objective is to evaluate an AI algorithm, specifically focusing on an Image Classification task as a case study. Real-world images are gathered and annotated with accurate labels, such as a cat or dog. A portion of these images is randomly selected to construct the training, validation, and test datasets based on a designated percentage. To assess the image classification algorithm (the subject in this case), it must be implemented on a computer system utilizing a specific programming framework, such as PyTorch or TensorFlow.

During the evaluation process, the test data is provided to the algorithm, which generates an output. This output is then compared to the ground truth labels. In Case Two, the ground truth can be considered as a test oracle. Additionally, measurements are collected for each run of the evaluation process.



The fundamental evaluation process in Case Two bears a resemblance to that of the baseline case. However, there exist three notable distinctions. Firstly, an EC in Case Two differs. The presence of a dataset labeled with ground truth constitutes a vital aspect of an EC. 
A dataset labeled with the ground truth represents a specific instance of a problem or task. 
Regrettably, in Case Two, it is impractical to define a problem or task mathematically. Hence, diverse problem or task instances
are devised in an ad-hoc manner, for example, selecting images randomly to form training, validation, and test datasets based on a predetermined percentage. This approach may introduce biases. 



Secondly, both the subject and the algorithms must be implemented on a computer, prompting the introduction of another concept called "a support system" to elucidate this facet, which may assume diverse manifestations in alternative evaluation scenarios. The support system represents an additional essential constituent of an EC. Therefore, in Case Two, we introduce two novel concepts: "the support system" and "the subject instantiation."  Figure~\ref{ec-definition} shows a hierarchy definition of an EC.

Thirdly, in Case Two, upon feeding the algorithm with the test data, it generates an output that is then compared with the ground truth, also known as the test oracle. Apart from measurements, there exist other forms of activities in the evaluation process, namely testing, as expounded upon in Section~\ref{testing}. Hence, in this instance, we modify the essence and definition of evaluation, as delineated in Section~\ref{essence_of_an_evaluation}, as follows. The essence of an evaluation lies in " conducting deliberate experiments where EECs are applied to a diverse range of subjects, resulting in the establishment of equivalent EMs. Subsequently, we can effectively evaluate the subjects by measuring and/or testing the equivalent EMs."
We formally define evaluation as "an experiment that applies EECs to diverse subjects and establishes equivalent EMs, enabling the measurement and/or testing of these equivalent EMs, the inference of the subjects' impact, and the subsequent judgment of them."



\textbf{Drug and policy evaluations}

The Third Case (Case Three) and the Fourth Case (Case Four) exhibit similarities, as they involve evaluating a drug and evaluating a policy aimed at addressing drug addiction within a community. In these cases, the subject refers to a specific drug or a policy aimed at addressing drug addiction intervention, while the support system encompasses the participants targeted by these interventions.

When comparing Cases Three and Four with Cases One and Two, we have made three observations. Firstly, the specific problem or task instances could be defined in a literal manner. Case Three focuses on the cure of a specific illness, whereas Case Four aims to address drug addiction and improve the overall well-being of individuals within a designated community. Regrettably, we currently lack a mathematical understanding of these problems or tasks, making it challenging to provide a mathematical definition of the abstract problem or task. Nevertheless, even if the problems can only be defined in a literal sense, having detailed and comprehensive definitions of problem or task instances, as well as a profound comprehension of the interrelationships between different problem or task instances (from biological or social perspectives), proves advantageous for the purposes of knowledge reuse and sharing. Furthermore, it is anticipated that in the future, we will strive to gain a deeper understanding of the connections between various diseases or social issues, potentially through mathematical means.

The second observation revolves around the distinctions of the support systems found in Cases Two, Three, and Four, specifically referring to the substantial variability in conditions within the target participants. Evaluations commonly utilize a methodology known as randomized controlled trials (RCT), which serves to eliminate confounding. 



In practical application, Randomized Controlled Trials (RCTs) are widely recognized as the gold standard for conducting evaluations. In an RCT, subjects and support systems in the treatment group and control group are randomly assigned. This random assignment helps to minimize confounding variables that may arise from differences in support systems. Additionally, the allocation of participants to either the treatment or control group is kept concealed from the evaluator and relevant stakeholders. 




Lastly, when it comes to Cases Three and Four, the third component of an EC often lacks a mathematical-form algorithm, although it relies on a scientifically valid mechanism that includes biological, social, or psychological interactions. That is the right reason for us proposing the term "algorithm-like mechanism," which we have explained before.  Moreover, Case Four presents a significantly more intricate situation compared to Case Three, as the instantiation of the algorithm-like mechanism encounters challenges in maintaining consistent quality due to diverse factors, such as different attitudes towards interventions and varying levels of communication skills that hold considerable importance.

In summarizing, across various evaluation scenarios in different disciplines, such as Cases One, Two, Three, and Four, it is possible to develop a comprehensive conceptual framework
that can be universally applied.

\subsubsection{The establishment of EECs}\label{EEC}


To lay the groundwork for the formalization of an EC, it is essential to establish a clear understanding of some crucial notations. The notations $E'$, $S$, and $U$ represent three crucial components. Specifically, $E'$ represents the problem or task space. $S$ represents the support system space. Finally, $U$ represents the subject space. $e'$, $s$, $u$ is an element of $E'$, $S$, and $U$ respectively. We note  $e'\in E'$, $s\in S$, $u\in U$.


In addition to the aforementioned notations, we also define several other fundamental notations. For each problem or task, $e'_i\in E'$, there is a set of problem or task instances noted as $E_i$. For all problems or tasks in $E'$, there is a collection of a set of problem or task instances, which we noted as $SE=(e'_i, E_i)$. We use the division notation $SE/E'$ to denote $E$. $E$ can be defined as the union of all $E_i$.

We introduce the notation $SA'$ to represent the algorithm-like mechanism space. This space, denoted as $SA'$, consists of a set of algorithm-like mechanisms that are associated with each problem or task instance. Specifically, for a given problem or task $e'_i$ in the problem or task space $E'$, and for each instance $e_{ij}$ in the corresponding instance space $E_i$, we define a set of algorithm-like mechanisms as $SA'=(e'_i, e_{ij}, A'_{ij})$.

To represent the algorithm-like mechanism space $SA'$ in relation to the problem or task space $E'$ and the problem or task instance space $E$, we use the division notation $SA'/E'/E$ to denote $A'$. $A'$ is a union of all $A'_{ij}$.



We introduce the notation $SA$ to represent the instantiations of the algorithm-like mechanism space. This space, denoted as $SA$, consists of a set of instantiations of algorithm-like mechanisms that are associated with each problem or task instance, algorithm-like mechanism, and support system. Specifically, for a given problem or task $e'_i$ in the problem or task space $E'$, for each instance $e_{ij}$ in the corresponding instance space $E_i$, for each algorithm-like mechanism $a'_{ijk}$ in the algorithm-like mechanism space $A'_{ij}$, and for each support system $s_l$ in the support system space $S$, we define a set of instantiations of algorithm-like mechanisms as $SA=(e'_i, e_{ij}, a'_{ijk}, s_l, A_{ijkl})$.

To represent the instantiations of the algorithm-like mechanism space $SA$ in relation to the problem or task space $E'$, the problem or task instance space $E$, the algorithm-like mechanism space $A'$, and the support system space $S$, we use the division notation $SA/E'/E/A'/S$ to denote $A$. $A$ is a union of all $A_{ijkl}$.

By introducing these notations, we establish a comprehensive framework that allows us to delineate the various components of an EC and their respective roles. This formalization enhances our understanding of the key components and their relationships within the EC framework.

We can formalize an EC as $C=E'\times E\times A'\times A\times S$.

In the realm of EC spaces, the concept of EECs plays a significant role. Two EC spaces, denoted as $C_1$ and $C_2$, are considered to be EECs if and only if there exists a bijection, denoted as $\beta$, between the two spaces: $\beta:C_1\mapsto C_2;\beta^{-1}:C_2\mapsto C_1$. 
This equivalence is denoted as $C_1\sim C_2$.


\subsubsection{LEEC and evaluation standard}~\label{LEEC}


In certain cases, establishing EECs can be a challenging task. It proves to be an arduous or unattainable task to ensure the equivalence of two ECs at all levels. However, it is crucial to ensure the relative comparability of evaluation outcomes. 




In this section, we propose the concept of the least equivalent evaluation conditions (LEEC)  as the foundation to attack this challenge.
In the event where we are unable to guarantee or establish the equivalence of two ECs at all levels in the hierarchy, we propose a minimum requirement of ensuring uniformity in the most essential components of the two ECs, which we refer to as the least equivalent evaluation condition (LEEC). 

We propose the establishment of LEEC at the levels of the first and second top components of ECs. 
Firstly, the first and second top components of an EC serve as the foundation upon which the other two lower-level components are derived. Therefore, they provide the most primitive components when setting ECs.  
Secondly, to enable effective comparison of evaluation outcomes, it is crucial to establish equivalence between the first high-level components of two ECs. If the first high-level components are not equivalent, the evaluation outcomes cannot be compared reliably. Thirdly, relying solely on the first component may not provide enough specificity and certainty. To address this, in addition to the equivalence of the first high-level component, it is necessary for two ECs to possess two sets of definite and solvable problem or task instances that are equivalent.


In certain situations, it is possible to relax the requirement of strict equivalency between the second high-level component if the scale of the problem or task instances can be defined. In such cases, we can consider a scenario where two ECs have the same problem or task but differ in the scales of problem or task instance as LEECs.


We formally define LEECs as follows:

For two ECs, denoted as $C_1=E'_1\times E_1\times A'_1\times A_1\times S_1$ and $C_2=E'_2\times E_2\times A'_2\times A_2\times S_2$, if there is equivalence between their first two subspaces ($E'$ and $E$), that is, if and only if there is a bijection, denoted as $\hat{\beta}$, between $E'_1\times E_1$ and $E'_2\times E_2$, we can establish that they are LEECs, denoted as $C_1\approx C_2$. In other words, $\hat{\beta}$ is a function mapping from $E'_1\times E_1$ to $E'_2\times E_2$, and its inverse function $\hat{\beta}^{-1}$, maps from $E'_2\times E_2$ to $E'_1\times E_1$, denoted as $\hat{\beta}:E'_1\times E_1\mapsto E'_2\times E_2;\hat{\beta}^{-1}:E'_2\times E_2\mapsto E'_1\times E_1$.




To effectively define the least equivalent EC (LEEC), it is crucial to identify the most governing component of ECs that must exhibit equivalency. This component, known as the \textit{evaluation standard}, plays a crucial role in defining the LEEC.  By establishing and adhering to this evaluation standard, we can ensure that the evaluation outcomes are relatively comparable.

An evaluation standard should embody the characteristics that are solvable,  definite, and equivalent (abbreviated as SDE).  First, it should be amenable to a solvable framework, employing specific mechanisms. These mechanisms could encompass mathematical steps executed in a mechanical fashion~\cite{denning1985science} or incorporate biological, social, and psychological mechanisms and interactions. It is noteworthy that algorithms can be regarded as specific applications of mathematical steps executed in a mechanical manner. If an evaluation standard does not lend itself to a solvable framework, it becomes essentially meaningless.   Second, it should possess definiteness, whereby there exists a unanimous understanding among evaluators without any uncertainty. Third, it should exhibit equivalence across multiple evaluators, ensuring consistency among their assessments.

We propose the establishment of an evaluation standard at the level of the definition of an individual problem or task instance. There are several valid reasons for this approach.

Firstly, the first and second high-level components define LEEC, and they are the cornerstone of the evaluation conditions. These two high-level components serve as the foundation upon which the other two low-level components are derived.

\begin{figure}
	\centering
		\includegraphics[scale=.47]{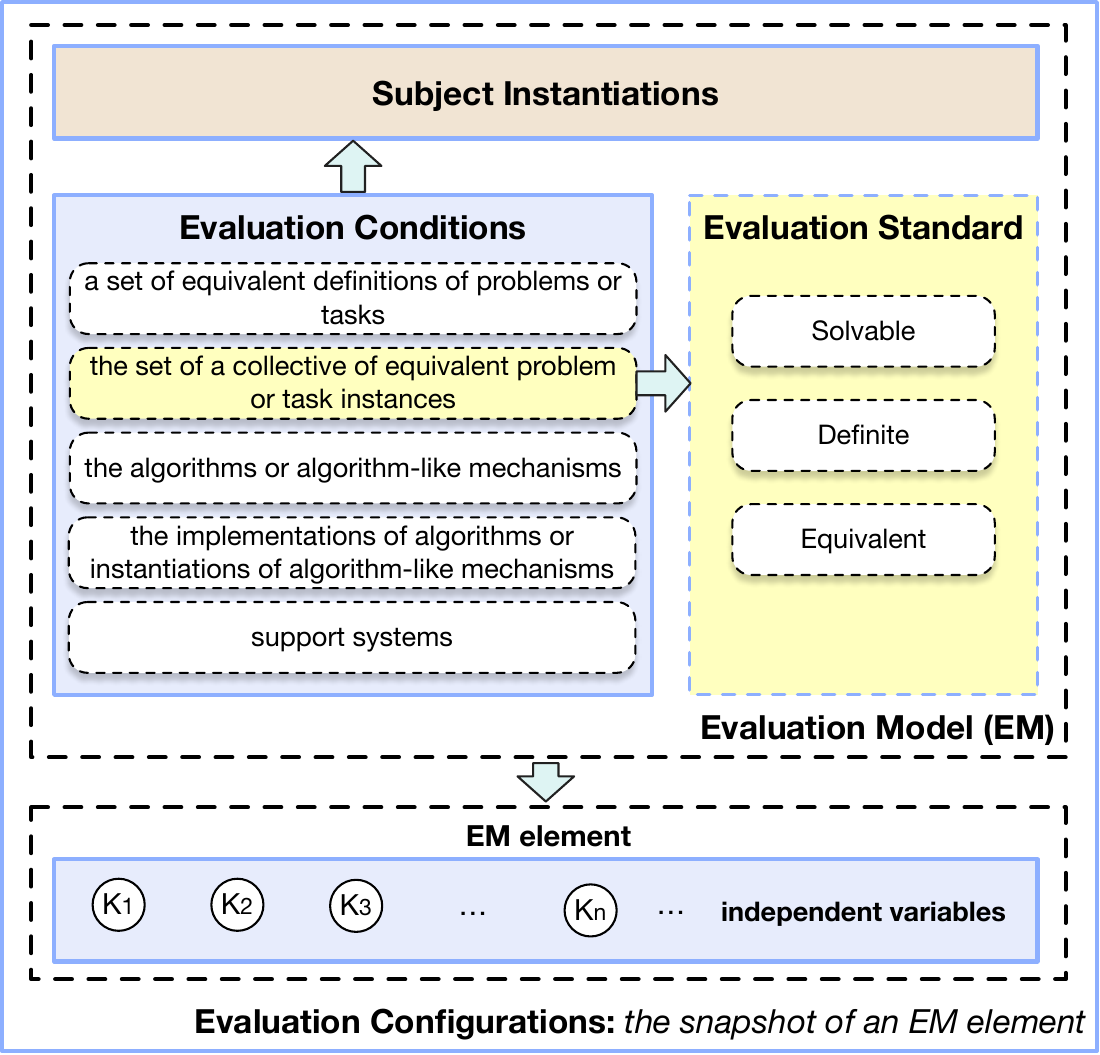}
	\caption{The relationships among evaluation configurations, evaluation conditions, and evaluation standards.}	\label{evaluation_config_condition_standard}
\end{figure}

Secondly, the first high-level component, which pertains to the definition of a problem or task, is not definite and specific, as it may encompass a population of different instances. To be precise, an equivalent, definite, and solvable problem or task instance qualifies as "an evaluation standard." This definition serves as the basis for conducting evaluations on the subjects. Figure~\ref{evaluation_config_condition_standard} shows the relationships among the evaluation standard, evaluation conditions, and evaluation configuration. 

We discuss the subtle differences between LEECs and evaluation standards. 
LEEC is defined on the first and the second high-level components, while the evaluation standard is only defined at the second high-level component, and they are closely related but with distinct implications. Their shared objective is to guarantee the comparability of evaluation outcomes. The aim of LEEC is to ensure the least equivalence between two ECs, while the evaluation standard is to provide the most governing component of an EC that ensures the comparability of the evaluation outcomes. Second, LEEC implies a state space, while an evaluation standard is an evaluation criterion. 





\subsubsection{The establishment of an REM}

Based on the notations defined in Section~\ref{EEC}, we formalize an EM as $M=C\times U=E'\times E\times A'\times A\times S\times U$. 

An element of an EM, denoted as $m\in M$, can be expressed as $m=(c,u)=(e',e,a',a,s,u)$. Here, $e'\in E'$ represents a given problem or task, $e\in E$ represents a specific instance of the problem or task, $a'\in A'$ represents a particular algorithm-like mechanism, $a\in A$ represents an instantiation of the algorithm-like mechanism, $s\in S$ represents a support system, and $u\in U$ represents a subject instantiation.

In addition, we will define how to establish an REM. We assume that an EM element, denoted as $m$, $\in M$,   is made of $n'$ independent variables, rewritten as $m=(k_1,\cdots,k_{n'}) =(e',e,a',a,s,u) $. We note $M=K_1\times\cdots\times K_n= C\times U=E'\times E\times A'\times A\times S\times U$. Please bear in mind that the number of variables, $n$, is greater than $n'$.


For each EM element, represented as $m=(k_1,\cdots,k_{n'})$, we follow a specific methodology in which only one independent variable at a time, from the set $k_1, k_2, ..., k_{n'}$, is allowed to vary while keeping the remaining variables constant. This controlled experimentation approach is referred to as an REM, as defined in Section~\ref{BEM}.




We define the evaluation cost of an EM or ES as the costs of constructing, traversing, and assessing its corresponding REM, where $\Vert M\Vert$ stands for the capacity of an EM or ES and $\mu$ stands for a constant coefficient:\\
\centerline{$\mathrm{cost}(M)=\mu\Vert M\Vert=\mu\Vert K_1\Vert\times\Vert K_2\Vert\times\cdots\times\Vert K_n\Vert$}



\subsection{Universal evaluation methodology in complex scenarios}~\label{universal_evaluation_methodology}

\begin{figure}
	\centering
		\includegraphics[scale=.5]{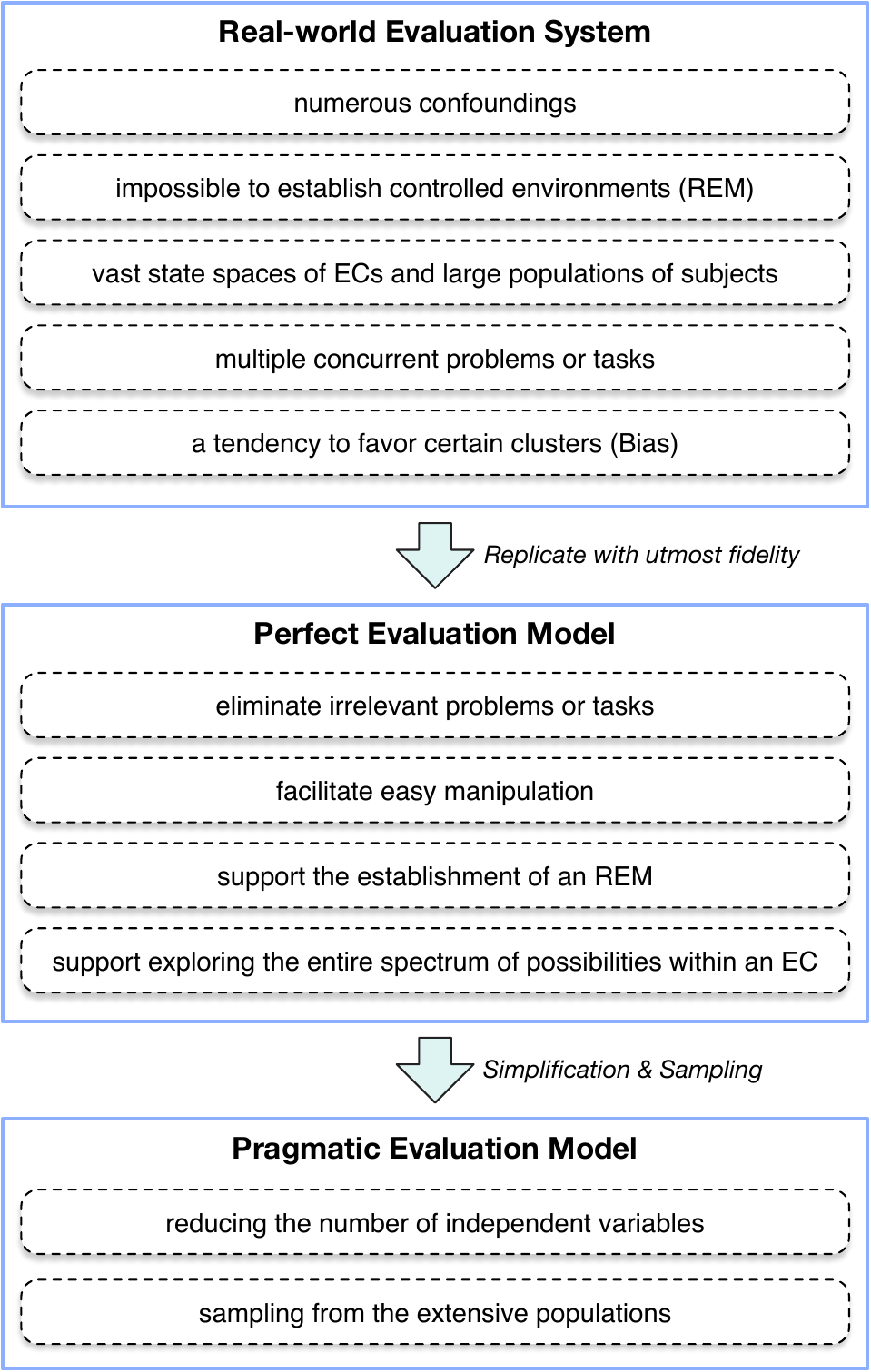}
	\caption{Universal evaluation methodology in complex scenarios.}	\label{evaluation-methodology-complex}
\end{figure}

From the revealed essence of the evaluation, it seems that performing an evaluation is straightforward. Unfortunately, in reality, there are evaluation scenarios with different levels of complexity. Figure~\ref{evaluation-methodology-complex} presents a universal evaluation methodology in complex scenarios. 

We refer to the entire population of real-world systems that are used to evaluate specific subjects as the \textit{real-world ES}.
Assuming no safety concerns are present, the real-world ES serves as a prime candidate for the assessment of the subjects of interest.  Unfortunately, there are five significant obstacles to consider when assessing diverse subjects within a real-world ES.  

Firstly, the presence of numerous confounding in the real-world ES poses a considerable challenge. It is often difficult, if not impossible, to completely eliminate these confounding. They can complicate the evaluation process by introducing variables that make it challenging to isolate the effects of different independent variables.


Secondly, manipulating the real-world ES is a formidable task, making it virtually impossible to establish controlled environments, known as REM,  for evaluating subjects. Additionally, the interconnected nature of subjects, support systems, and other components of ECs further complicates the establishment of an REM.


Thirdly, the vast state spaces of ECs and the large populations of subjects result in high evaluation costs. The sheer scale of these systems makes it expensive and time-consuming to thoroughly performing assessment and analysis.


Fourthly, within the real-world ES, multiple concurrent problems or tasks may be taking place simultaneously, which may not directly align with the subject being assessed. This further adds complexity to the evaluation process and introduces confounding.


Lastly, it is important to acknowledge that the real-world ES, regardless of the nature of its problems or tasks, tends to exhibit a bias towards certain clusters. This bias can manifest through problem or task instances, algorithm-like mechanisms, and their instantiations. However, this bias towards specific groups can limit our ability to fully explore and understand the entire range of possibilities available to us.

When examining a particular quantity of interest in an evaluation experiment, we define its \textit{accuracy} as the ratio between the quantity of the EM and the corresponding quantity of the real-world ES.   A higher ratio, closer to 100\%, indicates a better accuracy. On this basis,  we can define the \textit{accuracy of an EM} as a collection of accuracies of the quantities represented within the EM.  

In our research, we propose the concept of a "perfect EM" that aims to replicate the real-world ES with the highest level of fidelity, achieving a remarkable accuracy of 100\%. In theory, a perfect EM would possess several characteristics that enhance the evaluation of subjects within the EC framework.

Firstly, it would eliminate irrelevant problems or tasks that may be used to establish ECs, ensuring that the evaluation focuses on specific and directly applicable contexts. This targeted approach would enhance the relevance and applicability of the evaluation process.


Secondly, a perfect EM would facilitate easy manipulation, allowing for the free and artificial configuration of different evaluation settings. This flexibility would enable researchers to explore various scenarios and assess subjects under a range of conditions, enhancing the depth and breadth of the evaluation process.


Thirdly, a perfect EM would support the establishment of an REM, effectively eliminating confounding. By isolating and controlling variables of interest, researchers could gain more accurate insights into the impact of specific factors on the subjects being evaluated.



Furthermore, a perfect EM would have the capability to thoroughly explore and understand the entire spectrum of possibilities within an EC. This would include problem or task instances, algorithms-like mechanisms, and their instantiations. By encompassing this comprehensive range, researchers could gain deeper insights into the behavior and performance of subjects within the EC framework.

However, it is important to note that achieving a truly "perfect EM" may be challenging, if not impossible. The real-world ES is complex and dynamic, and replicating it with absolute fidelity is a monumental task. While we can strive to create more accurate and representative evaluation environments, it is crucial to recognize the inherent limitations and constraints that exist in the real world.

Nevertheless, by considering the concept of a perfect EM and its accompanying characteristics, we can strive to improve the evaluation of subjects within the EC framework and enhance our understanding of their performance within real-world contexts.




 The characteristics of the perfect EM, such as encompassing large populations of problem or task instances, algorithm-like mechanisms, instantiations, and support systems, as well as a vast number of independent variables, can lead to significant evaluation costs. However, to address this challenge, it is important to propose a pragmatic EM that simplifies the perfect EM in two key ways.

Firstly, to reduce the evaluation costs associated with a large number of independent variables, it is crucial to identify and focus on the variables that have a significant impact on the evaluation outcomes. By identifying and prioritizing these variables, researchers can streamline the evaluation process and allocate resources more efficiently. Negligible variables that have a minimal effect can be excluded or controlled for, reducing complexity and costs. It is worth emphasizing that the simplification involved in creating a pragmatic EM will inevitably lead to a decrease in the accuracy of the evaluation model.

Secondly, sampling techniques can be employed to manage the extensive populations of problem or task instances, such as algorithm-like mechanisms, instantiations, and support systems. Rather than evaluating every single possibility, researchers can select representative samples that capture the diversity and range of the population. This approach allows for a more manageable evaluation process while still maintaining a good level of coverage and representation.


\textit{In nature, a pragmatic EM is a subject equipped with a simplified and sampled EC.}
It can be considered as a sample of a perfect EM without taking into account any potential decrease in its accuracy. In order to measure the extent to which the statistics of a pragmatic EM can infer the parameters of a perfect EM,  we employ the criterion of confidence level and interval. 
The confidence level provides us with the probability that the estimated parameters of a perfect EM fall within a specific range of values. 
Meanwhile, the confidence intervals establish a range of values within which we can reasonably expect the true parameters of a perfect EM to fall. By utilizing these statistical measures, we can assess the degree of alignment between the statistics of a pragmatic EM and the parameters of a perfect EM. This allows us to gauge the effectiveness and validity of the pragmatic EMs.

By implementing these simplifications in a pragmatic EM, researchers can strike a balance between comprehensiveness and feasibility. The pragmatic EM allows for a more practical and efficient evaluation of subjects within the EC framework, mitigating the challenges posed by evaluation costs and the complexity of the perfect EM.

In representing different ECs, we use specific symbols. The symbol $C_r$ denotes the EC in a real-world ES (a real-world EC), which can be calculated as $E'_r\times E_r\times A'_r\times A_r\times S_r$. Similarly, the EC in a perfect EM (a perfect EC) is denoted by $C_p$, calculated as $E'_p\times E_p\times A'_p\times A_p\times S_p$. Lastly, the EC in a pragmatic EM (a pragmatic EC) is represented by $C_g$, calculated as $E'_g\times E_g\times A'_g\times A_g\times S_g$.

Likewise, we use symbols to denote a real-world ES ($M_r$), a perfect EM ($M_p$), and a pragmatic EM ($M_g$). These are represented as:

$M_r = E'_r\times E_r\times A'_r\times A_r\times S_r\times U_r$

$M_p = E'_p\times E_p\times A'_p\times A_p\times S_p\times U_p$

$M_g = E'_g\times E_g\times A'_g\times A_g\times S_g\times U_g$

These symbols help us distinguish and calculate the various components of ECs and EMs in different contexts.



\subsection{Fundamental issues in evaluatology}

This subsection presents three fundamental issues in Evaluatology. 

\subsubsection{Ensure transitivity of EMs}\label{transitivity}


The key to the effectiveness and efficiency of evaluations in different scenarios is to establish a series of EMs that ensure the transitivity of the primary characteristics.


In Section~\ref{universal_evaluation_methodology}, we have effectively examined and described a real-world ES, along with its corresponding EM, which we call a perfect EM, and their interconnections.

The real-world ES system presents three notable obstacles. Firstly, the presence of numerous confounding creates a challenge as they cannot be completely eliminated.  Secondly, the existence of multiple irrelevant problems or tasks adds another layer of complexity. Lastly, regardless of the nature of the problems or tasks involved, there is a tendency for the system to exhibit bias towards certain clusters exhibited by problem or task instances, algorithmic mechanisms, and their instantiations.

To overcome these obstacles, a perfect EM, which possesses several key characteristics, is proposed. Firstly, the model eliminates any irrelevant problems or tasks that may be used to derive evaluation standards for assessing different subjects.  Secondly, the model enables the free setting of an REM. Thirdly, the model allows for a comprehensive exploration and understanding of the entire spectrum of possibilities in terms of problem or task instances, algorithmic mechanisms, and their instantiations.




Compared to a real-world ES $M_r=E'_r\times E_r\times A'_r\times A_r\times S_r\times U_r$, a perfect EM eliminates its irrelevant problems or tasks, represented by $M_p=E'_p\times E_p\times A'_p\times A_p\times S_p\times U_p$. Transforming a real-world ES into a perfect EM should ensure the transitivity of the following characteristics. 

$E'_p\subset E'_r$.

$SE_p/E'_p\supset SE_r/E'_p$.

$SA'_p/E'_p/E_p\supset SA'_r/E'_p/E_p$.

$SA_p/E'_p/E_p/A'_p/S_p\supset SA_r/E'_p/E_p/A'_p/S_p$.  


\begin{figure}
	\centering
		\includegraphics[scale=.43]{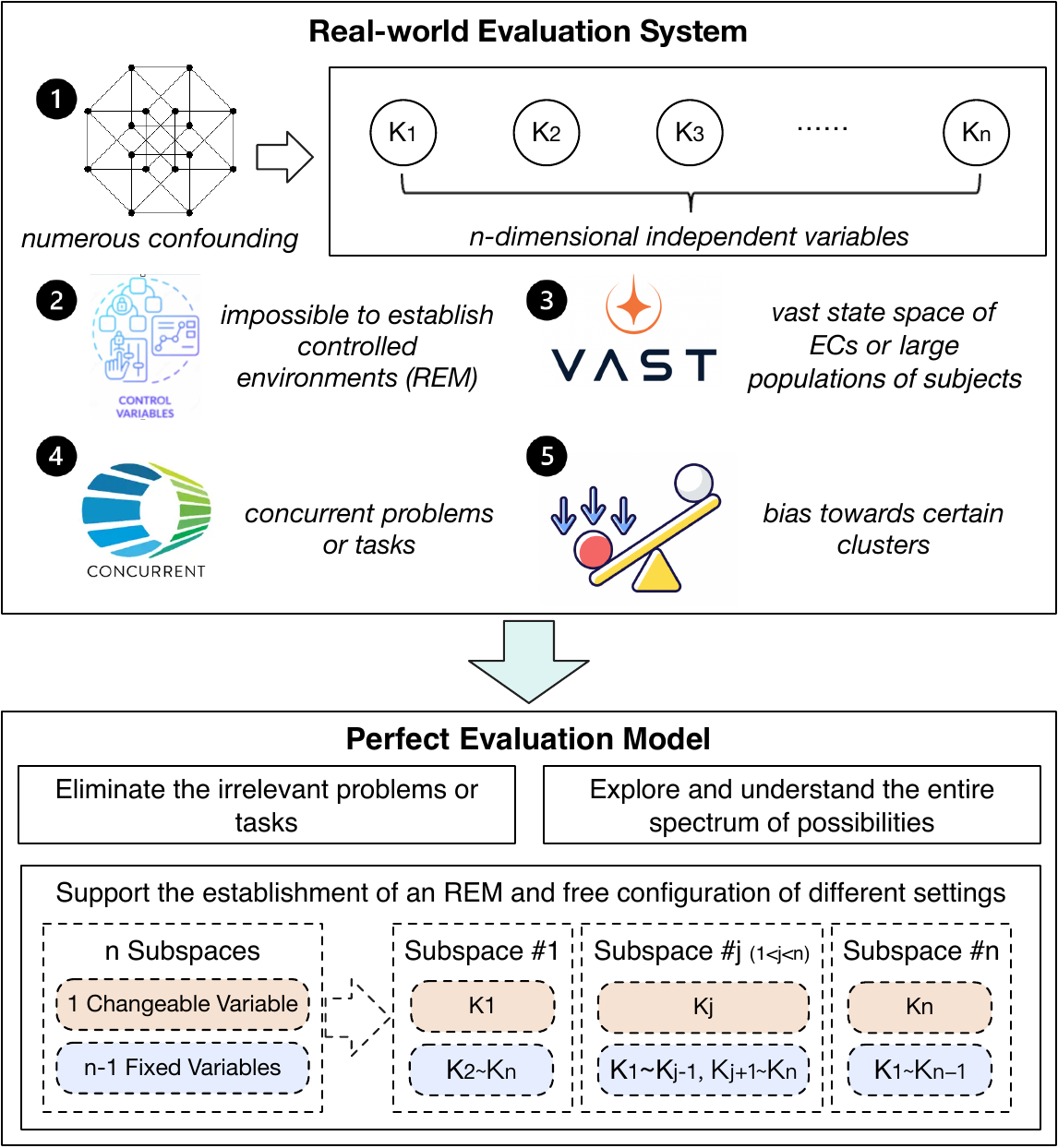}
	\caption{A perfect EM resembles a real-world ES.}
	\label{5.3.1}
\end{figure}

The perfect EM encompasses large populations of problem or task instances, algorithm-like mechanisms, their instantiations, and support systems. Also, it entails a vast number of independent variables. 

To overcome this difficulty, it becomes essential to propose a pragmatic EM that simplifies the perfect EM in two ways: (1) reducing the number of independent variables and (2) sampling from the extensive populations of support systems, problem or task instances, algorithm-like mechanisms, and their instantiations.

A pragmatic EM adopts a sampling approach on the perfect EM, resulting in a smaller space to work with. Additionally, the pragmatic EM streamlines the independent variables within the perfect EM by excluding those that have minimal impact. To formalize this process, we introduce the notation $s(\cdot)$ to represent the sampling function.

For each element $m_g$ in the sampled space $M_g$, which is a subset of the perfect EM $M_p$, we denote the corresponding element in $M_p$ as $m_{pg}$. When transforming a perfect EM into a pragmatic EM, it is essential to maintain the transitivity of the following characteristics:

$M_g = s(M_p)$: The sampled space $M_g$ is obtained through the application of the sampling function $s$ on the perfect EM $M_p$.

$M_g \subset M_p$: The sampled space $M_g$ is a subset of the perfect EM $M_p$.

$m_g = (k_1, \cdots, k_{n'}) \in M_g$: Each element $m_g$ in the sampled space $M_g$ consists of a set of independent variables $(k_1, \cdots, k_{n'})$.

$m_{pg} = (k_1, \cdots, k_{n''}) \in M_p, n' \leq n''$: The corresponding element $m_{pg}$ in the perfect EM $M_p$ consists of a set of independent variables $(k_1, \cdots, k_{n''})$. $n''$ is greater than $n'$, ensuring that the corresponding element in the perfect EM includes at least as many independent variables as that of the element in the pragmatic EM.

\subsubsection{Perform cost-efficient evaluation with controlled discrepancies}

By disregarding the accuracy of an EM, conducting evaluations solely through the establishment of an REM within a perfect EM may indeed result in maximum confidence. However, this approach also comes with a significant drawback - the exorbitant cost it entails. The process of creating an REM within a perfect EM can be prohibitively expensive, making it impractical for many organizations. Therefore, it is crucial to strike a balance between ensuring the discrepancy threshold of the evaluation outcomes and managing the associated costs when implementing evaluation processes.

When creating a pragmatic EM from a perfect EM, the discrepancy threshold $\epsilon$, which is a discrepancy limit that can be tolerated in an evaluation scenario, holds the potential to exert a profound influence on the evaluation results and, in certain instances, it would give rise to grave concerns, particularly in the context of safety-critical tasks where failure could lead to detrimental side effects such as harm, loss of life, or significant environmental damage. So, after thoroughly understanding the stakeholders' evaluation requirements, a risk function $\gamma(\cdot)$ could be predefined. When the stakes are high, and there is a greater risk associated with the evaluation outcomes, it becomes imperative to have a lower discrepancy threshold between the evaluation outcomes of a pragmatic EM and a perfect EM. This is because the potential consequences of making a wrong decision or drawing inaccurate conclusions become more significant.

In Section~\ref{transitivity}, we use the notation $s(\cdot)$ to represent the sampling function. In creating a pragmatic EM from a perfect EM, the accuracy of EM decreases. We use the notation $m(\cdot)$ to represent this modeling process. The accuracy of an EM is decided by $m(\cdot)$. We use the notation $e(\cdot)$ to represent the process of ensuring different equivalency levels of EC.

We introduce a discrepancy function of the evaluation outcomes $\mathrm{disc}(\cdot)$ between $M_g$ and $M_p$.  
When the discrepancy is $0$, it indicates that $M_g$ and $M_p$ are equivalent.


The discrepancy function of the evaluation outcomes $\mathrm{disc}(\cdot)$ between $M_g$ and $M_p$ is defined as follows. In the formulation, $\rho(\cdot)$ is a measurement function, and $v(\cdot)$ is a value function. Besides, we define the evaluation cost as the product of a constant $\mu$ and the space capacity of $M_g$. This cost factor allows us to incorporate the resource constraints and practical considerations associated with the evaluation process.

$$
\left\{
\begin{aligned}
&\mathrm{discrepancy\ threshold\ \epsilon}=\mathrm{\gamma(\cdot)}, \\
&\mathrm{disc(M_g, M_P)}=\mathrm{disc}(v(\rho(c(m(s(M_p))))),v(\rho((M_p))),\\
&\mathrm{accuracy(M_g)=m(s(M_p))},\\
&\mathrm{cost}=\mu(\Vert M_g\Vert).
\end{aligned}
\right.
$$

Based on the above formulation, the evaluation issue of balancing evaluation cost and the discrepancies in the evaluation outcomes can be framed as an optimization problem. The objective is to minimize the evaluation cost, represented by $\mathrm{cost}$ while ensuring the discrepancies in the evaluation outcomes, denoted as $\mathrm{disc}(M_g, M_p)$, do not exceed a predefined discrepancy threshold $\epsilon$.


The optimization problem can be formulated as follows:\\
\centerline{$\arg\min\limits\mathrm{cost(M_g)}\text{ subject to }(\mathrm{disc}(M_g, M_p)<\epsilon)$}.

\begin{figure}
	\centering
		\includegraphics[scale=.43]{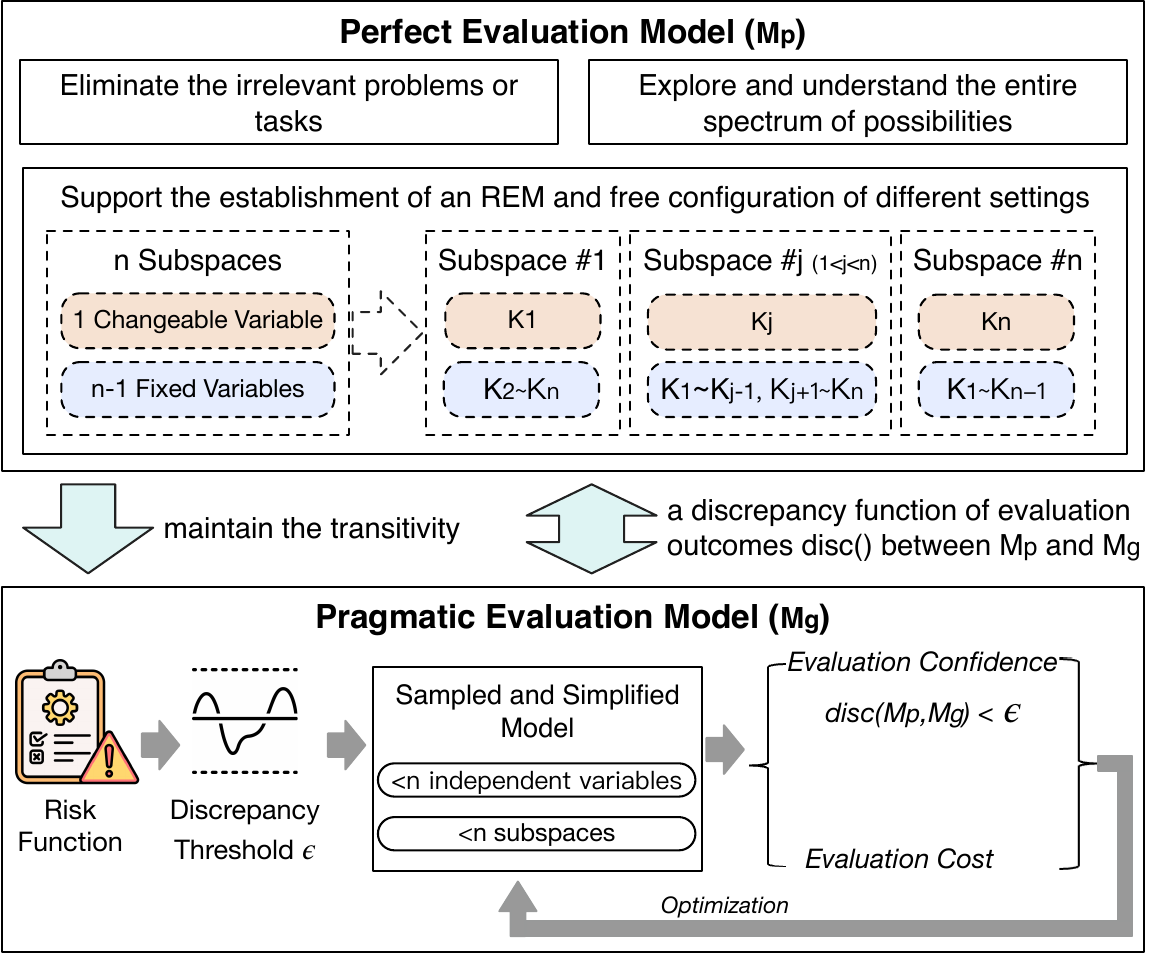}
	\caption{Proposing a Pragmatic EM based on the Evaluation Risk Function.}
	\label{5.3.2}
\end{figure}

\subsubsection{Ensure evaluation traceability}\label{evaluation_traceability_issue}


According to the third axiom of evaluation, for a well-defined subject, the divergence in the evaluation outcomes can be attributed to disparities in ECs, thereby establishing evaluation traceability. 



\begin{figure}
	\centering
		\includegraphics[scale=.43]{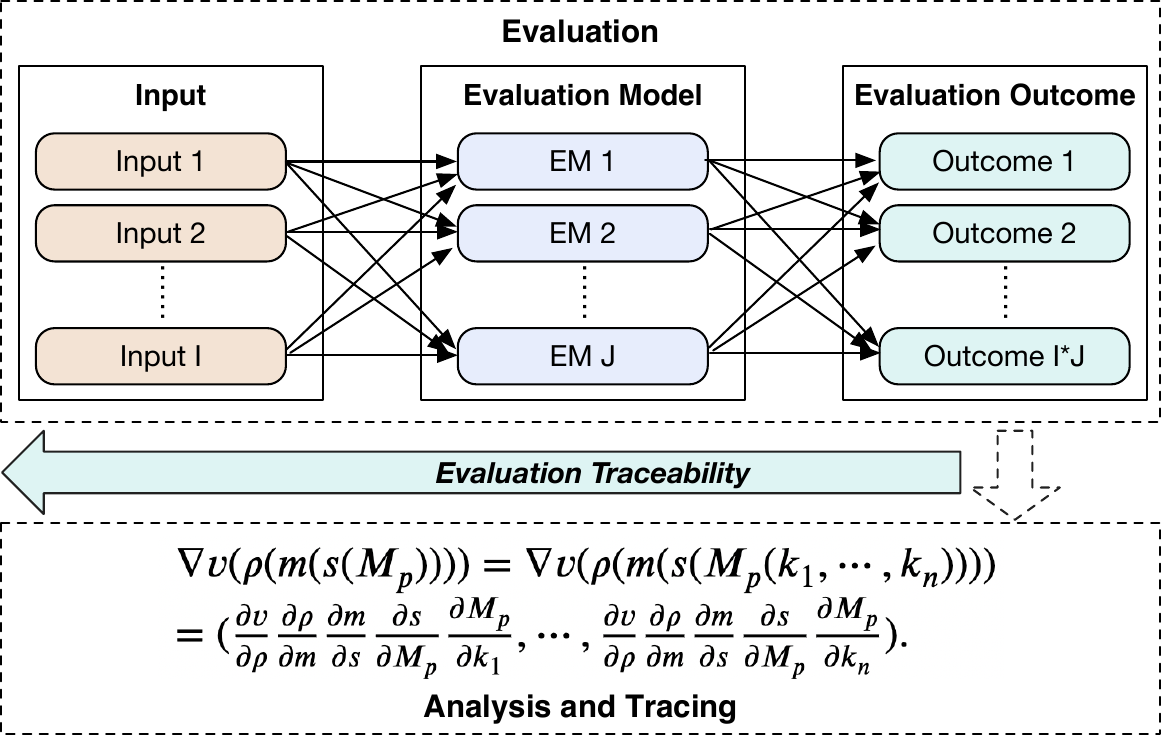}
	\caption{Evaluation traceability.}
	\label{5.3.3}
\end{figure}

Conceptually, traceability asks for a quantified mapping between the differences in the input and output of the value function $v(\cdot)$ decided by the evaluation community and the measurement function $\rho(\cdot)$ through the evaluation process, described by the mathematical model we formulated above, and the model in the following formulation is a perfect EM. We discover that this concept aligns well with the mathematical notation of the gradients of a function, which gives the rate of changes in the output for each input variable. 
Figure~\ref{5.3.3} shows how to ensure the evaluation traceability.
In the context of evaluation, the gradient of evaluation outcomes can be written as follows, which is a matrix or tensor:\\
$\nabla v(\rho (c(m(s(M_p)))))=\nabla v(\rho (c(m(s(M_p(k_1,\cdots,k_n)))))$\\$=( \frac{\partial v}{\partial\rho} \frac{\partial \rho}{\partial c} \frac{\partial c}{\partial m}  \frac{\partial m}{\partial s} \frac{\partial s}{\partial M_p} \frac{\partial M_p}{\partial k_1},\cdots,\frac{\partial v}{\partial \rho} \frac{\partial \rho}{\partial c} \frac{\partial c}{\partial m}  \frac{\partial m}{\partial s} \frac{\partial s}{\partial M_p} \frac{\partial M_p}{\partial k_n})$.\\
The closed-form mathematical expression is not always available for various EC components in evaluation. Nevertheless, we can follow the method of acquiring gradients in numerical methods by creating perturbations in the ECs for various input variables and observing the differences in the composite evaluation outcomes, thus approximating the gradients.

\subsubsection{Connect and correlate evaluation standards across diverse disciplines}

While the constituents comprising ECs may differ across distinct evaluation scenarios, a governing fundamental component of ECs emerges as the evaluation standard. The shared qualities of evaluation standards, as denoted by the SDE characteristics, indicate the potential for establishing a correlation between evaluation standards across diverse disciplines.

The evaluation standard serves as a fundamental pillar within any evaluation model. By establishing connections between evaluation standards across various disciplines, we have the potential to construct a comprehensive framework encompassing evaluation issues in all fields. This holistic framework, known as the grand unified theory of evaluatology, allows for a thorough exploration of evaluation-related matters.

Before defining an evaluation standard, it is essential to understand the nature of the stakeholders' primary problems or tasks.
In the rest of this article, a problem or task refers explicitly to a computational problem. Make sure to distinguish between a problem and a problem instance: A problem is an infinite collection of problem instances, each of which is a problem with concrete configurations, which we have elaborated in Section~\ref{EC_definition}.

Computational complexity theory provides the basis for understanding the nature of primary problems. For example, complexity classes --- that are defined by bounding the time or space used by the algorithm --- can be used to understand the problem's nature~\cite{johnson1990catalog}. Computability theory~\cite{cooper2017computability} seeks a more general question about all possible algorithms that could be used to solve the same problem. That is to say, the computability theory provides a viable solution to answer whether a problem is solvable. In understanding the problem instance, the theory of analysis of algorithms~\cite{sedgewick1996introduction} can be used to analyze the amount of resources needed by a particular algorithm to solve a problem instance. 

While the computational complexity and other theories mentioned above lay the foundation, the formulation of evaluation standards introduces novel concerns. The first challenge lies in addressing scenarios where articulating a mathematical model explicitly becomes an insurmountable task. Regrettably, this circumstance is not uncommon, as numerous problems defy expression through a mathematical model at present.

\begin{figure*}
	\centering
		\includegraphics[scale=.47]{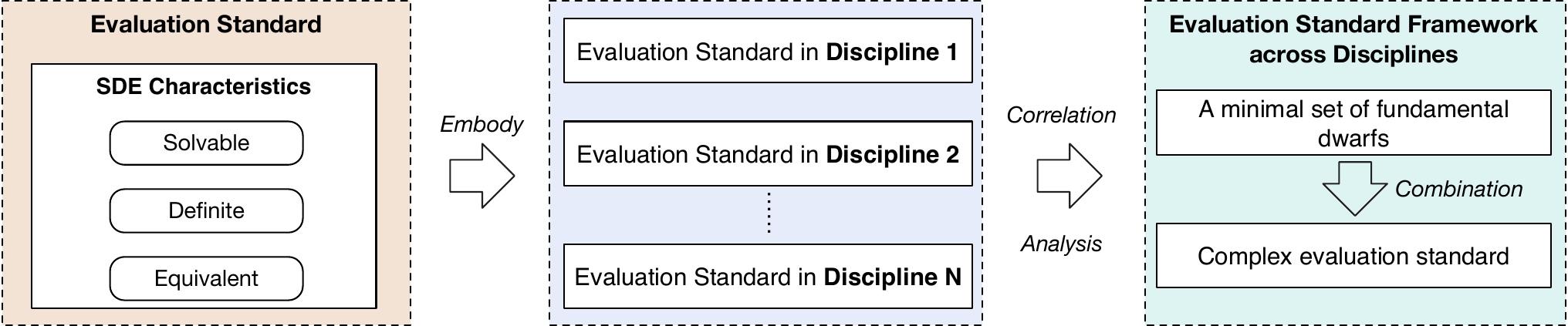}
	\caption{Correlating evaluation standards across diverse disciplines.}
	\label{5.3.4}
\end{figure*}


To ensure the definitiveness and equivalency of the evaluation standard, it is imperative to establish a rigorous problem space definition and a problem instance space definition, which provides the quantitative foundation for the comparability and traceability of different EMs.

The representativeness of the evaluation standard is a crucial aspect that warrants discussion. Understanding the composition of problems is crucial in identifying the problem or task that best represents the whole. For instance, across various scientific and engineering disciplines, problems often exhibit a hierarchical structure, where a significant problem can be broken down into several smaller problems commonly referred to as "dwarfs"~\cite{asanovic2006landscape}. This pattern can be considered one of the foundational structures within problem domains. Gaining profound insights into the structural aspects of problems proves immensely valuable when assessing complex and multifaceted subjects.

Figure~\ref{5.3.4} shows how to correlate evaluation standards across diverse disciplines.
In an optimal scenario, we can identify evaluation standards that embody the SDE characteristics across various disciplines. Ultimately, we can establish connections and correlations between evaluation standards from different fields, giving rise to the grand unified theory of evaluatology. The objective of this theory is to present a hierarchical framework of evaluation standards. Within this framework, we can identify a minimal set of fundamental evaluation standard "dwarfs" along with their respective variations. Complex evaluation standards are formed by combining two or more of these evaluation standard dwarfs and their variants. This hierarchical structure of evaluation standards will greatly facilitate the reuse and sharing of knowledge.

\section{Benchmarkology: the engineering of evaluation}~\label{benchmarkology} 

This section unveils the core essence of a benchmark and introduces the benchmark-based engineering of evaluation, which we call benchmarkology. Furthermore, we provide guidelines and workflows within the realm of benchmarkology.


\subsection{The essence of a benchmark}\label{essence_benchmark}

Benchmarks are extensively employed across various disciplines, albeit lacking a formal definition. Based on the science of evaluation, we propose a precise delineation of a benchmark as \textit{a simplified and sampled EC, specifically a pragmatic EC, that ensures different levels of equivalency, ranging from LEECs to EECs.}

Within the framework of this definition, a benchmark comprises three essential constituents. The first constituent is the \textit{stakeholder's evaluation requirements}, which encompass various factors. These include the risk function, which evaluates the potential risks associated with the benchmark. Additionally, the discrepancy threshold, which determines the acceptable level of deviation in evaluation outcomes, is considered. The evaluation confidence and interval play a crucial role in predicting the parameter of a perfect EM, while the accuracy of EM determines the level of accuracy achieved by the benchmark. Lastly, the evaluation cost of EM is taken into account, and the resources required for conducting the evaluation are assessed. By considering these elements, the benchmark can effectively address the evaluation requirements of stakeholders.

The second constituent of the benchmark framework is the \textit{EC configuration and mechanisms}. This includes several elements crucial for the benchmark's effectiveness. Firstly, it involves defining the set of problems or tasks that the stakeholders face when addressing them. Additionally, it encompasses the set of equivalent problem or task instances, which helps ensure specificity in the evaluation process. The benchmark also considers algorithm-like mechanisms and their instantiations, which play a significant role in solving the defined problems or tasks. The support systems, which provide necessary resources and environments, are also taken into account.

Moreover, the benchmark provides the means to eliminate confounding variables that may affect the evaluation outcomes. Finally, the benchmark provides the mechanism to ensure varying levels of EC equivalency, determining the extent to which different benchmark instances can be considered equivalent.  By considering these EC configurations and mechanisms, the benchmark can provide a comprehensive and standardized approach to evaluating problems or tasks.



The third constituent is the \textit{metrics and reference}, including the measurement and testing procedures, the definitions of quantities,  the value function, composite evaluation metrics,  the reference subject, and the reference evaluation outcomes.

\begin{figure*} 
	\centering
		\includegraphics[scale=.47]{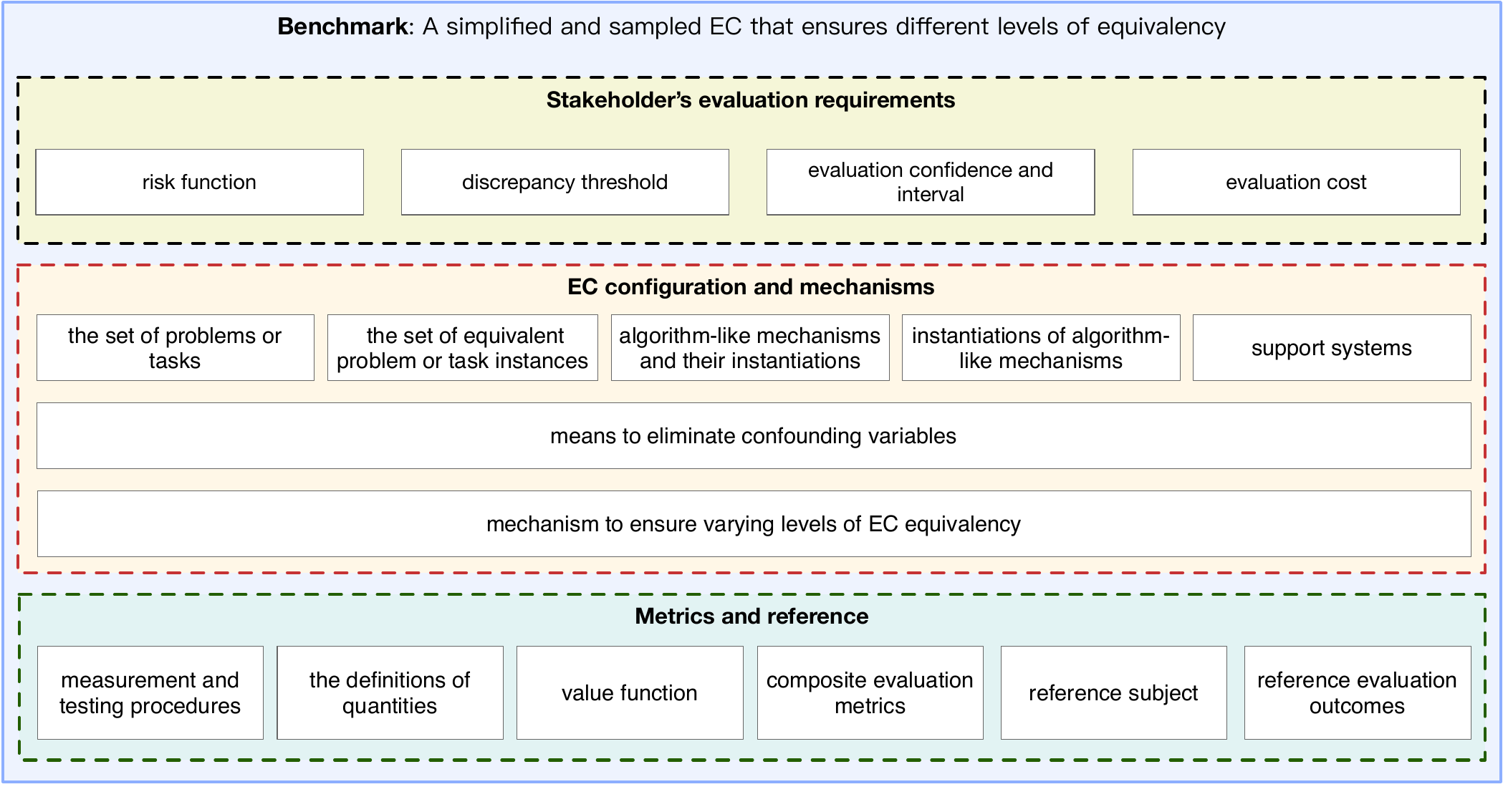}
	\caption{A benchmark comprises three essential constituents.}\label{benchmark_essence_constituents}
\end{figure*}

In the subsequent sections of this article, we will refer to these three constituents as the complete constituents of a benchmark. Figure~\ref{benchmark_essence_constituents}  shows the three essential constituents of a benchmark. 

\subsection{The goal of benchmarkology}

As expounded upon in Section~\ref{evaluationonology}, the science and engineering of evaluation, known as evaluatology, aims to apply the EECs to various subjects and establish an REM. A benchmark can be viewed as a simplified and sampled EC, specifically a pragmatic EC, that ensures different levels of equivalency, ranging from LEECs to EECs. In this context, a benchmark-based approach to the evaluation problem is considered a feasible engineering methodology, given the widespread use of benchmarks across various disciplines. Consequently, we propose the formal definition of \textit{benchmarkology} as an engineering discipline concerned with the quantitative assessment of diverse subjects using benchmarks. 



Undoubtedly, the theory of evaluatology serves as the foundation for benchmarkology. Nevertheless, benchmarkology has its unique objective – to furnish guiding principles and engineering evaluation methodologies.

 \subsection{The principles in building benchmarks.}

In Section~\ref{evaluationonology}, we have extensively discussed the fundamental axioms of evaluatology. Nonetheless, this particular subsection delves deeper into the principles that underpin the creation of benchmarks derived from these four evaluation axioms.



\textbf{The first principle focuses on the validity of metrics within a benchmark.} According to the First Axiom of Evaluation, also known as the Axiom of the Essence of Composite Evaluation Metric, there are three criteria to determine the validity of metrics in benchmarks. If a metric does not meet these criteria, it is considered invalid. The three criteria are as follows:
\begin{enumerate}
    \item The metric should be a base quantity.
    \item The metric should represent another quantity that has inherent physical significance.
    \item The metric should be a composite evaluation metric that is explicitly defined by a value function.
\end{enumerate}

~\textbf{The second principle pertains to the comprehensiveness of the configurations of a valid benchmark.} According to the Second Axiom of Evaluation, known as the Axiom of True Evaluation Outcomes, when a well-defined subject is applied to a well-defined EC, its evaluation outcomes possess true values. A well-defined EC should reveal all its well-defined components. Each component within the EC plays a critical role in determining the evaluation outcomes. When the components of the EC are not well-defined, uncertainty is introduced into the evaluation process. Without clear and specific components, the evaluation outcomes become unpredictable and lack reliability. 


The evaluation outcome using a benchmark should reveal the complete evaluation configurations. Unfortunately, many contemporary benchmarks, both in terms of state-of-the-art and state-of-the-practice, have failed to disclose their comprehensive evaluation configurations fully.

Some benchmarks, like the widely-used CPU benchmark SPECCPU, may omit certain constituents or their components, e.g., the support system. In such cases, it becomes essential to clearly define the conditions under which the simplification is made, ensuring that the benchmark can still provide meaningful and valid results. By providing these detailed evaluation configurations, we can ensure that the benchmark remains a reliable tool for evaluation purposes.





~\textbf{The third principle centers around the concept of benchmark traceability.} In accordance with the Third Axiom of Evaluation, also known as the Axiom of Evaluation Traceability, it is crucial to establish benchmark traceability to enable the comparison of different benchmarks. This means that it is a top priority to trace the discrepancies in evaluation outcomes back to variations in the configurations of different benchmarks.

~\textbf{The fourth principle encompasses the validity of the comparison ratio of two evaluations obtained using the benchmark}. It emphasizes the importance of understanding the comparison ratio in relation to the evaluation confidence and the evaluation interval of its accuracy.


Comparison is a common practice using benchmarks, and the comparison ratio or speedup is often a key metric of interest. When we compare two evaluations, we use the benchmarks as a means to infer the parameters of the real-world ES. In this context, the accuracy of the comparison ratio corresponds to an evaluation interval with a specific confidence level. So, the validity of the comparison ratio of two evaluations obtained using the benchmark must be understood in relation to the evaluation confidence and evaluation interval of its accuracy. 


To illustrate this principle, let's consider a specific example where we compare two subjects, Subjects A and B, and obtain a speedup ratio (either greater than or less than 1).

In a simulated CPU scenario, the reported speedup ratio is 1.3, indicating an improvement in performance for Subject A compared to Subject B. However, the accuracy of the speedup ratio has a confidence interval of [0.7, 1.9] with a 90\% confidence level. This means that the true value of the speedup ratio could be any value within this interval, including 1.6.

If the accuracy of the speedup ratio is indeed 1.6, the actual speedup in the real-world ES would be 1.3 divided by 1.6, resulting in a value of 0.8. This indicates a degraded performance for Subject A in comparison to Subject B, reported on the real-world ES, which is contradicted by an improvement in performance reported on a simulated system (an EM).

It is crucial to note that relying solely on the reported speedup ratios without considering the confidence level and the confidence interval of its accuracy can lead to misleading interpretations and decisions. To ensure the validity of evaluation outcomes, it is essential to take into account the confidence interval and the confidence level associated with the speedup ratio.

\subsection{The universal methodology in benchmarkology}\label{methodology_benchmarkology}

\begin{figure} 
	\centering
		\includegraphics[scale=.52]{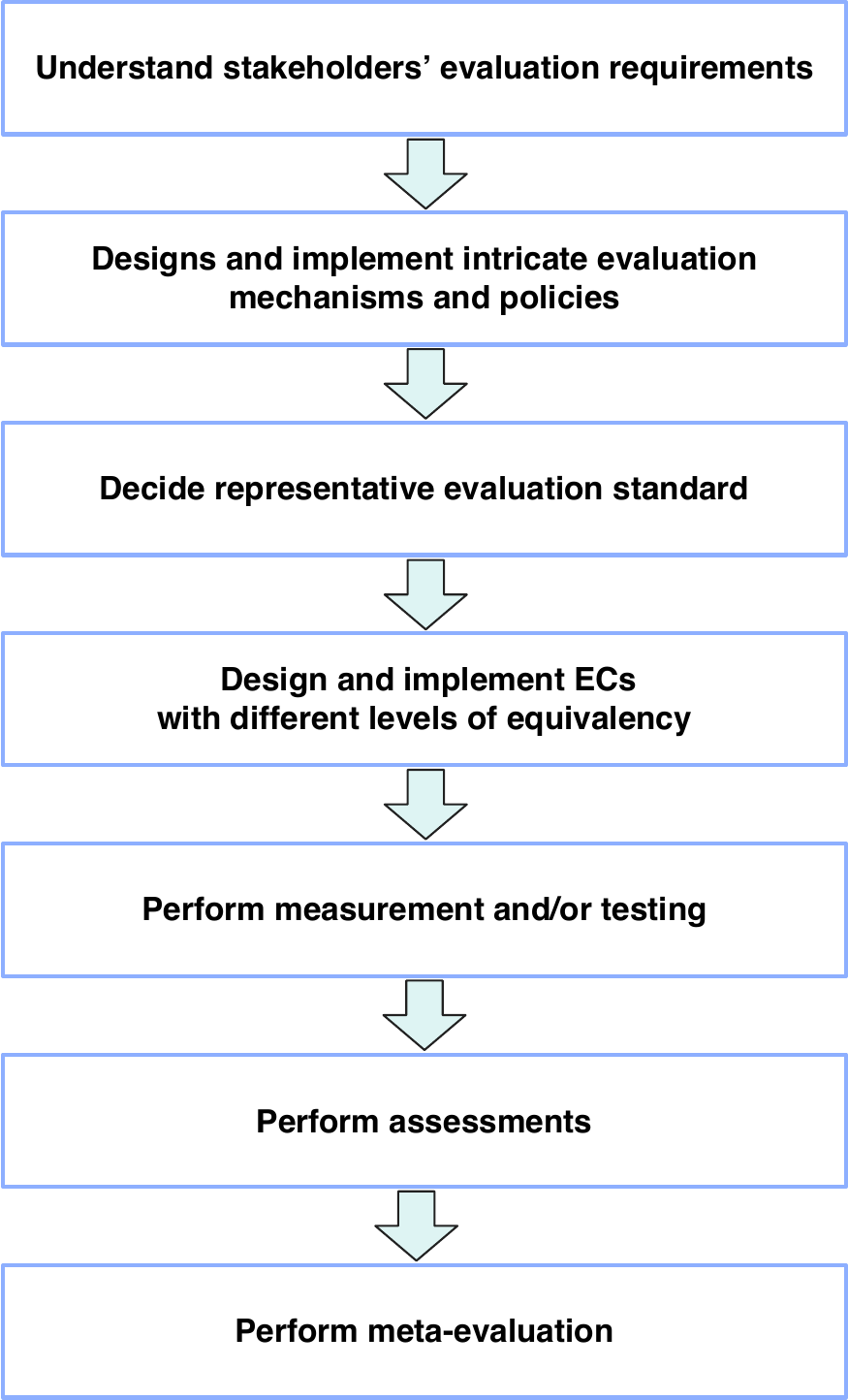}
	\caption{The universal methodology in benchmarkology.}
	\label{benchmark_universal}
\end{figure}

The aforementioned principles offer valuable insights into the fundamental components of a benchmarkology workflow, as shown in Fig.~\ref{benchmark_universal}.

\subsubsection{Understand stakeholders' evaluation requirements}



During the initial phase, it is essential for evaluators to gain a comprehensive understanding of the stakeholders' evaluation requirements, the first constituent of the benchmark, which we discussed in detail in Section~\ref{essence_benchmark}.  This crucial step allows for the alignment of these requirements with the overall purpose of the evaluation.


During this phase, a thorough examination of the evaluation risk function, the discrepancy threshold of evaluation outcomes, the accuracy, and the evaluation cost should be conducted. Additionally, for quantities or variables of interest, it is crucial to establish their confidence level and confidence interval when using the benchmark. This quantification allows for an assessment of how effectively the benchmark can infer or predict the parameters of a perfect EM.

Unfortunately, in state-of-the-art or state-of-the-practice benchmarks, the importance of this phase is often overlooked. There are two possible reasons for this oversight.

Firstly, in certain evaluation scenarios, the discrepancy in evaluation outcomes, whether intermediate or large, may not have significant consequences. However, it is crucial to note that this is not the case in scenarios involving safety-critical, mission-critical, and business-critical applications. In these situations, even minor deviations can have severe impacts on the overall outcome.

Secondly, the benchmark process itself is an engineering practice that emphasizes iterative and refined operation. As a result, it implicitly incorporates some procedures of this phase.

Therefore, it is imperative to recognize the significance of understanding stakeholders' evaluation requirements, particularly in scenarios where the stakes are high and any discrepancy from expected outcomes can have critical implications.




\subsubsection{Design and implement  intricate evaluation mechanisms and policies}~\label{Phase_Two}

This phase plays a crucial role, particularly in complex evaluation scenarios, and can be quite costly. Its primary aim is to provide a solid foundation for generating a benchmark.




The real-world ES reflects the complexities and nuances of actual evaluation environments.
 In this phase, the evaluator takes on the crucial task of building 
 and investigating the real-world ES. The methodology discussed in Section~\ref{universal_evaluation_methodology} can serve as a helpful guide for this process.

During the investigation, evaluators need to consider several aspects carefully. One aspect involves identifying and eliminating irrelevant problems or tasks that may not be applicable to the assessment of the subjects under evaluation. This ensures that the evaluation focuses on relevant and meaningful aspects of the subject. Another important consideration is that evaluators must recognize any constraints or factors that may impact the evaluation process. This understanding lays the groundwork for creating a perfect EM that explores all the possibilities. A perfect EM serves as an ideal evaluation model, and the design and implementation of this model are also key focus in this phase. We have discussed its main concerns in Section~\ref{universal_evaluation_methodology}.

Additionally, in this phase, two key policies and their accompanying procedures are of utmost importance. Firstly, the modeling policy and procedure guide the process of transforming the real-world ES into an EM, striking a balance between accuracy and cost. This involves capturing the essential elements and characteristics of the real-world ES in the model while ensuring that the modeling process is efficient and cost-effective.

Secondly, the sampling policy and procedure play a vital role in transitioning from a perfect EC  to a pragmatic EC. This transition aims to save on evaluation costs while still maintaining a high level of confidence in the evaluation results. The sampling policy and procedure guide the selection of representative samples from the perfect EC, ensuring that the pragmatic EC captures the essential aspects and characteristics of the perfect EC while being more practical and resource-efficient.

By following these policies and procedures, the evaluation process is adapted to real-world conditions and constraints. The modeling policy and procedure enable the creation of an EM that accurately represents the real-world ES, while the sampling policy and procedure ensure that the pragmatic EC reflects the essential elements of the perfect EC. This allows for a more effective and efficient evaluation process that balances accuracy, cost, and confidence.

Overall, with the facilitation of the study of real-world ES and the perfect EM, these modeling and sampling policies and procedures are essential in this phase to guide the modeling and sampling processes, ensuring that the evaluation process is well-suited to real-world conditions and constraints. 


\subsubsection{Decide representative evaluation standards}

The third phase is to decide the representative evaluation standards that guarantee the least EECs, ensuring the comparability of the evaluation outcomes.  The main objective of this phase is to carefully consider the relevant stakeholders involved and gain a deep understanding of their principal interests and concerns. This phase requires evaluators to identify and comprehend the primary problems or tasks that need to be addressed.

It is important to note that each stakeholder has a unique perspective, leading to subtle differences in the problems or tasks they face. Therefore, it is crucial for evaluators to recognize and take into account these varying perspectives, ensuring that the evaluation standards are comprehensive and reflective of the diverse interests of the stakeholders involved.

In Section~\ref{LEEC}, we have explored the concept of evaluation standards and how they are derived from the abstract problem or task at hand. Building upon this understanding, the evaluator's next step is to determine the specific evaluation standards by selecting representative instances of the primary problems or tasks faced by the stakeholders.

It is important to recognize that different stakeholders will have distinct evaluation standards. This is because their perspectives and priorities vary based on their unique roles and interests. As a result, it is essential for evaluators to consider these differences and tailor the evaluation standards accordingly to ensure that they capture the specific needs and concerns of each stakeholder involved.



It is true that previous evaluation and benchmark practices have often lacked consistent discussions on what qualifies as an evaluation standard. However, we believe that our universal definition of evaluation standards has the potential to be applicable across various evaluation scenarios in different disciplines. Our aim is to provide a comprehensive framework that can guide evaluators in establishing effective evaluation standards.

Furthermore, we acknowledge the importance of recognizing that a realistic benchmark can only capture a small sample of huge populations of instances that are derived from a large population of problems or tasks. Despite the challenges that may arise when explicitly stating the problem or task, we remain committed to adopting a systematic approach to our thinking. This allows us to navigate through such complexities and develop meaningful evaluation standards that align with the objectives of the evaluation process.


\subsubsection{Design and implement ECs with different levels of equivalency}

Based on the outputs from Phases Two and Three, the subsequent stage involves the design and implementation of ECs with different levels of equivalency.  This task varies from different subjects. In general, this phase needs to consider algorithm-like mechanisms and their instantiations, as well as the support systems.

Furthermore, in this phase, it is essential to address the levels of EC equivalency. This involves determining the extent to which different benchmark instances can be considered equivalent. It requires careful consideration of which components can be disregarded or simplified in order to streamline the benchmark process while maintaining its validity and reliability.

Additionally, the benchmark needs to establish mechanisms to eliminate confounding that may impact the evaluation outcomes. Confounding variables can introduce biases or distortions into the evaluation results, affecting their accuracy and reliability. One approach to address confounding is by employing our proposed REM methodology. This methodology provides a systematic framework to identify and eliminate confounding variables and ensure that the evaluation outcomes are not influenced by extraneous factors.

\subsubsection{Perform measurement and/or testing}

The fourth phase encompasses measurements and/or testing, guided by the principles and practices of the metrology and testing theory.  The measurement and testing process serves multiple purposes. Firstly, evaluators must determine which properties or quantities to measure, keeping in mind what base quantities and other quantities that carry physical meaning are.  

Furthermore, evaluators must also consider the cost of measurement and testing, ensuring that it aligns with budgetary constraints. It becomes imperative for them to make informed decisions on various aspects, such as how, when, and to what extent to perform testing, sampling, and measuring these properties or quantities. By doing so, they can effectively manage resources while still obtaining valuable data for their research.

\subsubsection{Perform assessments}

The fifth phase entails assessment, wherein the defining of a value function takes precedence. The rationale behind establishing a value function lies in the aim of encapsulating numerous quantities that surpass our capacity for recognition into a singular metric. This value function serves as a proposed function, mapping the target properties or quantities measured during the preceding phase to the evaluation outcomes in order to reflect the concerns or interests of the stakeholders. Given that stakeholders often possess varying concerns or interests, it is common to propose multiple value functions from different perspectives. Different communities may reach a consensus on how to define a value function. Generally, evaluators must engage in consultation with the stakeholders to define a composite evaluation metric in the form of a value function that truly encapsulates the stakeholders' concerns or interests. 


Subsequently, evaluators can compare the obtained evaluation outcomes to the reference evaluation outcomes, ultimately enabling judgment on different subjects, such as performance, value, merit, weaknesses, worth or significance, as well as positive or negative effects.


\subsubsection{Perform meta-evaluation}

The last phase involves conducting meta-evaluations. The evaluators are tasked with reviewing all the evaluation processes and determining whether the theory, data, or evidence produced can substantiate the conclusions drawn in the evaluations. In this phase of meta-evaluation, various perspectives are taken into account, including the dimensionality of measurement data~\cite{furr2011scale}, the reliability of measurement results, the validity of evaluation outcomes, the traceability of benchmarks, as well as the cost and cost-efficiency of the evaluation itself~\cite{rossi2018evaluation}. 

The "dimensionality" of measurement data refers to the number and nature of variables that are reflected in the assessment~\cite{furr2011scale}. The reliability of measurement results pertains to the extent to which the measured values accurately reflect the true values. The validity of the evaluation outcome denotes the degree to which the statistics of the benchmark can infer the parameter of the real-world evaluation setting or a perfect evaluation model in terms of the metrics of confidence level and confidence interval. 

Two approaches can be undertaken to manage benchmark traceability. Firstly, it is essential to develop a comprehensive mathematical model capable of capturing the influence of the discrepancies of the evaluation configurations on the evaluation results. This model will serve as a foundation for interpreting and comprehending the evaluation outcomes. 
Additionally, the community must engage in continual benchmark comparisons. Drawing inspiration from the calibration practices in the field of metrology, this consistent comparison and alignment of different benchmarks will ensure consistency and accuracy in evaluation procedures. 



Constrained by the limitations of the budget, the aforementioned six phases can be carried out iteratively, employing a trial-and-error methodology.

\section{Why it is essential to develop evaluatology?} ~\label{SubSection_Motivation} 


In this section, we will begin by conducting a comprehensive review of the state-of-the-art and state-of-the-practice evaluation methods and benchmarks. This review will provide us with a solid understanding of the current landscape and help identify areas where advancements are needed.

To illustrate the importance of advancing the science and engineering of evaluation, we will focus on the evaluation of CPU performance as a prime example. 
Meanwhile, we will critically reflect on the existing state-of-the-art and state-of-the-practice evaluation methods and benchmarks. This reflection will enable us to identify any limitations or gaps that need to be addressed for more accurate and meaningful evaluations.

Furthermore, we will explore the advantages that the field of evaluatology brings to the table. 
To ensure clarity and understanding, we will provide a concise summary of the distinctions between evaluation, measurement, and testing. 




To ensure consistency and alignment with our proposed universal terminology, we will utilize our established terminology framework when discussing state-of-the-art and state-of-the-practice evaluation and benchmark cases. 

\subsection{Evaluations across different academic fields}\label{related_work}

This subsection presents a concise overview of the cutting-edge evaluations in a range of academic disciplines, as well as the prevailing evaluation practices. 

\subsubsection{Observation study methodologies}


Observational study methodologies are widely used in the fields of business science, finance, and education. Even based on a random sample, an observational study still falls short of effectively revealing the cause-and-effect relationships.

\textbf{Evaluations in the field of business science:}

 Camp~\cite{camp1989benchmarking} defines benchmarking as "the search for those best practices that will lead to the superior performance of a company." Benchmarking consists of two primary steps~\cite{camp1989benchmarking}: (1) establishes operation targets based on industry best practices; (2) "a positive, proactive, structured process leads to changing operations and eventually achieving superior performance and competitive advantage." 
In the study conducted by Andersen et al.~\cite{andersen1999benchmarking}, the essence of benchmarking is summarized as the quest for knowledge and learning from others. 



\textbf{Evaluations in the fields of finance and education:}

In the fields of finance and education, indices are widely used as benchmarks to assess the overall performance of the individuals or systems under study. These indices are derived by calculating the weighted average of a selected group of individuals or systems ~\cite{fisher1966some}.

For example, stock market indices are widely used as benchmarks to assess the overall performance of the stock market in the field of finance. These indices are derived by calculating the weighted average of a selected group of representative stocks~\cite{fisher1966some}. Some well-known stock market indices include the Dow Jones Industrial Average, the S\&P 500, the NASDAQ Composite, the Shanghai Stock Exchange Composite Index, and the Hang Seng Index. Different indices employ varying calculation methods. The most common approach is the weighted average method, which determines the index value based on the weighted average of the constituent stock prices. Another method is the geometric mean method, which calculates the geometric average of the stock prices and adjusts it using a base period price. Typically, stock market indices are published at the close of each trading day. Some index providers offer real-time index data, enabling investors to stay informed about the latest market conditions.

The Brent benchmark is used to determine the price of Brent crude oil~\cite{SCHEITRUM2018462}. Brent crude oil is a type of light and low-sulfur crude oil produced from oil fields in the North Sea region. Due to its relatively stable supply and high quality, Brent crude oil has become a significant benchmark in the international oil market. Traders, investors, and industry participants worldwide reference the Brent benchmark to track and evaluate the price of Brent crude oil.

In the finance discipline, indexes or benchmarks serve as reference measurements or evaluation results. However, these practices often prioritize data collection and processing over building a solid evaluation theory framework.

\subsubsection{Experimental methodologies}

Experimental methodologies are widely used in the fields of social sciences, computer sciences, psychology, and medicine. 

\textbf{Evaluations in the field of social sciences:}

According to Rossi et al., ~\cite{rossi2018evaluation}, at the earliest, Thomas Hobbes and his contemporaries tried to "use numerical measures to assess social conditions and identify the cause of mortality, morbidity, and social disorganization in the discipline of social science."

Rossi et al.~~\cite{rossi2018evaluation} define program evaluation as the process of using social research methods to systematically assess programs aimed at "improving social conditions and our individual and collective well-being," with the goal of providing answers to the stakeholders.
Rossi et al.~~\cite{rossi2018evaluation} summarize the five domains of evaluation questions and methods that exhibit strong interplays: (1) the need for the programs, (2) program theory and design, (3) program process, (4) program impacts, and (5) program efficiency.

\textbf{Evaluations in the Field of Computer Science:}



The SPEC CPU benchmark suite, known as SPEC CPU~\cite{SPECCPU}, is widely recognized as the most renowned benchmark suite for CPU performance evaluation. Throughout its history, six versions of the SPEC CPU benchmark suite have been released, with the latest version being SPEC CPU2017, which can be found in Table~\ref{CPU_Evaluation_Conditions}. The SPEC CPU workloads cover a broad range of CPU-intensive tasks.

The performance evaluation metric used in SPECCPU is based on the execution time. The reported score of SPECCPU represents the ratio of its execution time compared to that of a reference machine. The specific details of a reference machine can be found in Table~\ref{CPU_Evaluation_Conditions}. To ensure the credibility of the results, the overall metrics are calculated as the geometric mean of each respective ratio. Each ratio is based on the median execution time from three runs or the slower of the two runs.

Dongarra et al.~\cite{2010The} proposed the LINPACK benchmark for evaluating high-performance computing (HPC) systems. The LINPACK Benchmark is designed to solve dense linear systems of equations of order n, represented by the equation $Ax=b$. It originated from the development of the LINPACK software package in the 1970s.

The LINPACK benchmark is commonly used to evaluate HPC systems, and the measurement metric is the number of floating-point operations per second (FLOPS). FLOPS represents the count of floating-point operations (FLOPs) performed by the solving algorithm of the LINPACK benchmark, which is calculated as ($2*n^{3}/3 + 2*n^{2}$) operations divided by the execution time of the benchmark. The LINPACK benchmark is also used by the TOP500 list of the world's most powerful supercomputers to rank their performance.

\begin{figure}
	\centering
    \includegraphics[scale=.5]{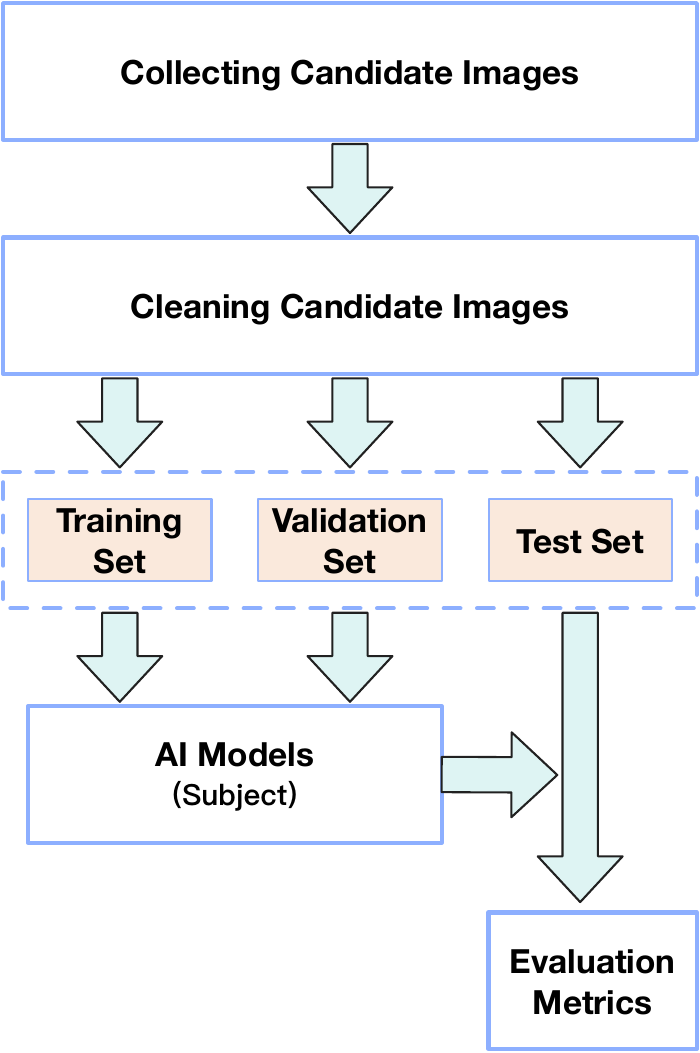}
	\caption{The ImageNet evaluation working process.}
	\label{imagenet_process}
\end{figure}

As shown in Figure~\ref{imagenet_process}, ImageNet is a significant benchmark in the field of computer vision, consisting of 14,197,122 high-resolution images manually annotated across 21,841 distinct categories, commonly known as ImageNet-21K~\cite{deng2009imagenet}. These categories encompass a wide range of objects, animals, and scenes. The ILSVRC (ImageNet Large Scale Visual Recognition Challenge) is an annual computer vision competition that focuses on a subset of ImageNet-21K called ImageNet-1K~\cite{russakovsky2015imagenet}. It aims to evaluate the performance of deep learning models in tasks such as image classification and object detection, providing specific task configurations and evaluation criteria. ImageNet-1K is primarily used for image classification tasks and consists of 1,281,167 training images, 50,000 validation images, and 100,000 test images. The evaluation metrics commonly used in ILSVRC include Top-1 accuracy, which measures the match between the predicted category and the true category of the image, and Top-5 accuracy, which indicates if the true category of the image is among the top five predicted categories by the model.


\textbf{Evaluations in the field of medicine:}

The evaluation in the field of medicine can be traced back to the early medical eras, although there are no documented records. A rigorous modern medical evaluation methodology and system were established as early as 1938~\cite{cavers1939food}. Clinical trials, with a history spanning over 250 years, are the primary and widely recognized method for medical evaluation. They are defined as experimental designs to evaluate the potential impact of medical interventions on human subjects~\cite{jenkins1991history}.

Currently, clinical trials based on experimental designs can be categorized into various types, including randomized trials, double-blind trials, prospective trials, and retrospective trials~\cite{sellers2022descriptive}.

\begin{figure}
	\centering
		\includegraphics[scale=.5]{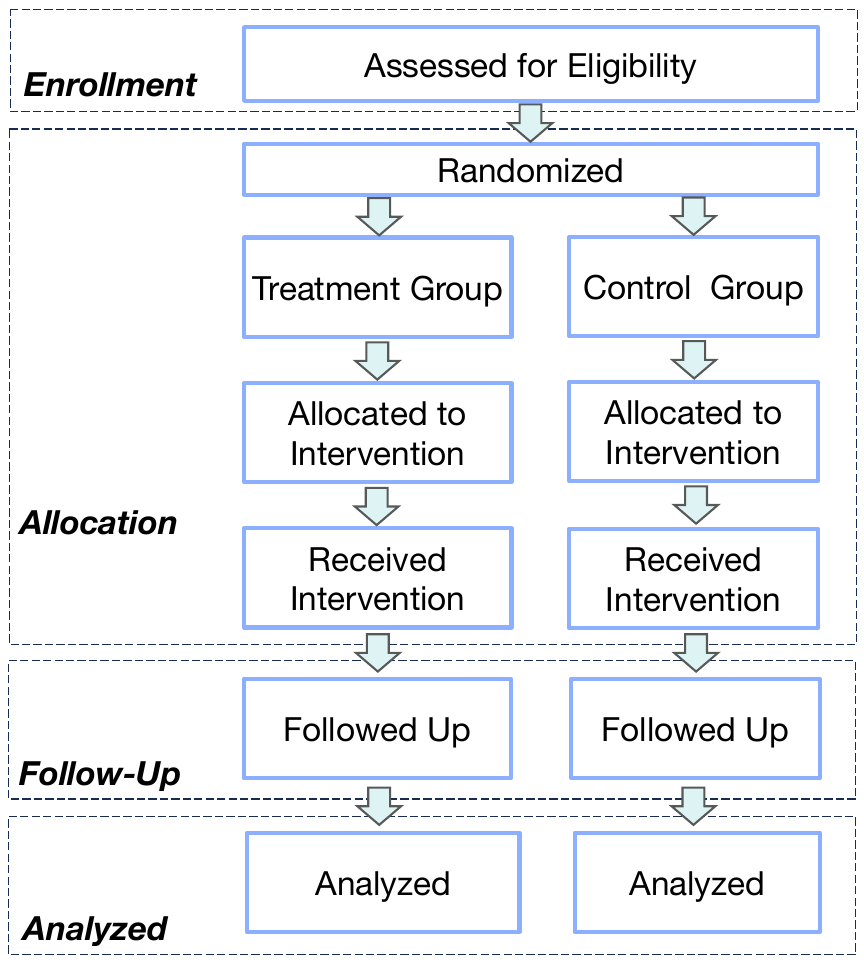}
	\caption{The randomized controlled trials (RCT) evaluation process~\cite{moher2001consort}}
	\label{RCT_process}
\end{figure}

As illustrated in Figure~\ref{RCT_process}, Randomized Controlled Trials (RCTs), considered the gold standard for medical evaluation, possess a rigorous and reliable theoretical framework~\cite{monti2018randomized}. However, their high time and financial costs limit their application. To compensate for the shortcomings of RCTs, emerging clinical evaluation methods, such as Real-World Data (RWD) assessment and digital clinical trials, have been proposed~\cite{inan2020digitizing,ramagopalan2020can}. These novel medical assessments are still in their early stages and have noticeable deficiencies in their theoretical foundations, such as lacking rigor and reliability.




\textbf{Evaluations in the field of psychology:}

In the field of psychology, social and personality psychologists often rely on scales, such as psychological inventories, tests, or questionnaires~\cite{furr2011scale}, to evaluate psychometric variables~\cite{furr2011scale}. These variables include attitudes, traits, self-concept, self-evaluation, beliefs, abilities, motivations, goals, social perceptions, and more~\cite{furr2011scale}.

It is important to note that cognitive biases, which are systematic patterns of deviation from norm or rationality in judgment~\cite{bias2015evolution}, may introduce distortions in self-report style evaluations.




\subsection{A case study on CPU benchmarks}\label{CPU_Benchmark_case_study}

Within this scenario, we assess the same CPU utilizing diverse CPU benchmarks, namely SPEC CPU2006~\cite{CPU2006}, PARSEC 3.0~\cite{bienia2008parsec}, and SPEC CPU2017~\cite{SPECCPU2017}, which are proposed by distinct entities employing diverse methodologies. 


\begin{table*}[]
\caption{The base quantity, value function, and the reference machine specified in different CPU benchmark suites in Section~\ref{SubSection_Motivation}.}
\scriptsize
\begin{tabular}{|c|l|l|l|l|}
\hline
\textbf{Benchmark} & \multicolumn{1}{c|}{\textbf{Category}} & \multicolumn{1}{c|}{\textbf{Base Quantity}}& \multicolumn{1}{c|}{\textbf{Value Function}}   &  \multicolumn{1}{c|}{\textbf{Reference Machine}} \\ \hline
\multirow{2}{*}{\begin{tabular}[c]{@{}l@{}}\\ \\ SPEC CPU2006\end{tabular}} & SPECspeed & Execution Time & \begin{tabular}[c]{@{}l@{}}$x_i=$ time on a reference machine / \\ time on the evaluation machine~\cite{SPECCPU2017}; $\overline{x}=\sqrt[n]{\prod \limits_{i}^n x_i}$\end{tabular}  & \multirow{2}{*}{\begin{tabular}[c]{@{}l@{}}\\the Ultra Enterprise 2 with 296 MHz\\ UltraSPARC II chips~\cite{CPU2006}\end{tabular}} \\ \cline{2-4}
 & SPECrate & Execution Time & \begin{tabular}[c]{@{}l@{}}$x_i=$ number of copies * (time on a reference \\ machine / time on the evaluation machine)~\cite{SPECCPU2017}; $\overline{x}=\sqrt[n]{\prod \limits_{i}^n x_i}$\end{tabular} &  \\ \hline
\multirow{2}{*}{\begin{tabular}[c]{@{}l@{}}\\ \\ SPEC CPU2017\end{tabular}} & SPECspeed & Execution Time & \begin{tabular}[c]{@{}l@{}}$x_i=$time on a reference machine /\\time on the evaluation machine~\cite{SPECCPU2017}; $\overline{x}=\sqrt[n]{\prod \limits_{i}^n x_i}$\end{tabular} & \multirow{2}{*}{\begin{tabular}[c]{@{}l@{}}\\the Sun Fire V490 with 2100 MHz\\ UltraSPARC-IV+ chips~\cite{SPECCPU2017}\end{tabular}} \\ \cline{2-4}
 & SPECrate & Execution Time & \begin{tabular}[c]{@{}l@{}}$x_i=$number of copies * (time on a reference \\ machine / time on the evaluation machine)~\cite{SPECCPU2017}; $\overline{x}=\sqrt[n]{\prod \limits_{i}^n x_i}$\end{tabular} &  \\ \hline
PARSEC &  & Execution Time & $x_i=$ time on the evaluation machine  & \multicolumn{1}{c|}{\diagbox{}{}} \\ \hline
\end{tabular}
\label{CPU_Evaluation_Conditions}
\end{table*}

\begin{table*}[]
\caption{The evaluation outcomes of the Intel Xeon Gold 5120T Processor using different CPU benchmarks show significant discrepancies. SPEC CPU2017 includes two sub-suites: SPECrate and SPECspeed. The results derived from SPECrate and SPECspeed are further categorized into two groups, known as floating-point (FP) and int.}
\scriptsize
\label{CPU_Benchmark_Evaluation}
\begin{tabular}{|l|l|l|l|l|c|l|}
\hline
\textbf{Benchmark}                                & \multicolumn{1}{c|}{\textbf{Support Platform}}                                                                                   & \multicolumn{1}{c|}{\textbf{Workload}} & \textbf{Time (s)} & \multicolumn{1}{c|}{\textbf{Score}}   & \multicolumn{1}{c|}{\textbf{Result}}     & \multicolumn{1}{c|}{\textbf{Score Meaning}}                                   \\ \hline
\multicolumn{1}{|c|}{}                            &                                                                                                                                  & 503.bwaves\_r                          & 1483              & 379.0             &                           &                                                                               \\ \cline{3-5}
\multicolumn{1}{|c|}{}                            &                                                                                                                                  & 508.namd\_r                            & 636               & 83.7     &                                  &                                                                               \\ \cline{3-5}
\multicolumn{1}{|c|}{}                            &                                                                                                                                  & 510.parest\_r                          & 1742              & 84.1      &                                 &                                                                               \\ \cline{3-5}
\multicolumn{1}{|c|}{}                            &                                                                                                                                  & 511.povray\_r                          & 1128              & 116.0      &                                  &                                                                               \\ \cline{3-5}
\multicolumn{1}{|c|}{}                            &                                                                                                                                  & 519.lbm\_r                             & 1420              & 41.6      &                                 &                                                                               \\ \cline{3-5}
\multicolumn{1}{|c|}{}                            &                                                                                                                                  & 521.wrf\_r                             & 1682              & 74.6     &                                  &                                                                               \\ \cline{3-5}
\multicolumn{1}{|c|}{}                            &                                                                                                                                  & 526.blender\_r                         & 787               & 108.0      &                                  &                                                                               \\ \cline{3-5}
\multicolumn{1}{|c|}{}                            &                                                                                                                                  & 527.cam4\_r                            & 998               & 98.2     &                                  &                                                                               \\ \cline{3-5}
\multicolumn{1}{|c|}{}                            &                                                                                                                                  & 538.imagick\_r                         & 1479              & 94.2    &                                   &                                                                               \\ \cline{3-5}
\multicolumn{1}{|c|}{}                            &                                                                                                                                  & 544.nab\_r                             & 893               & 106.0       &                                 &                                                                               \\ \cline{3-5}
\multicolumn{1}{|c|}{}                            &                                                                                                                                  & 549.fotonik3d\_r                       & 2001              & 109.0       &                                 &                                                                               \\ \cline{3-5}
\multicolumn{1}{|c|}{}                            &                                                                                                                                  & 554.roms\_r                            & 1429              & 62.3      &                                 &                                                                               \\ \cline{3-6}
\multicolumn{1}{|c|}{}                            &                                                                                                                                  & 500.perlbench\_r                       & 926               & 96.3      &   \multicolumn{1}{c|}{\multirow{-12}{*}{96.9 (FP)}}                             &                                                                               \\ \cline{3-5}
\multicolumn{1}{|c|}{}                            &                                                                                                                                  & 502.gcc\_r                             & 758               & 105.0       &                                 &                                                                               \\ \cline{3-5}
\multicolumn{1}{|c|}{}                            &                                                                                                                                  & 505.mcf\_r                             & 1059              & 85.5    &                                   &                                                                               \\ \cline{3-5}
\multicolumn{1}{|c|}{}                            &                                                                                                                                  & 520.omnetpp\_r                         & 1217              & 60.4      &                                 &                                                                               \\ \cline{3-5}
\multicolumn{1}{|c|}{}                            &                                                                                                                                  & 523.xalancbmk\_r                       & 786               & 75.3      &                                 &                                                                               \\ \cline{3-5}
\multicolumn{1}{|c|}{}                            &                                                                                                                                  & 525.x264\_r                            & 1179              & 83.2      &                                 &                                                                               \\ \cline{3-5}
\multicolumn{1}{|c|}{}                            &                                                                                                                                  & 531.deepsjeng\_r                       & 715               & 89.8       &                                &                                                                               \\ \cline{3-5}
\multicolumn{1}{|c|}{}                            &                                                                                                                                  & 541.leela\_r                           & 1197              & 77.5       &                                &                                                                               \\ \cline{3-5}
\multicolumn{1}{|c|}{}                            &                                                                                                                                  & 548.exchange2\_r                       & 1338              & 110.0        &                                &                                                                               \\ \cline{3-5}
\multirow{-22}{*}{SPECrate} & \multirow{-22}{*}{\begin{tabular}[c]{@{}l@{}}Unix\\ (AIX,\\  HP-UX,\\  Linux,\\  Mac OS X,\\  Solaris),\\  Windows~\cite{SPECCPU2017}\end{tabular}} & 557.xz\_r                              & 824               & 73.4            &    \multicolumn{1}{c|}{\multirow{-10}{*}{84.3 (INT)}}                      & \multirow{-22}{*}{\begin{tabular}[c]{@{}l@{}}Higher scores mean \\ that more work is \\ done per unit of time~\cite{SPECCPU2017}\end{tabular}}                \\ \hline
                                                  &                                                                                                                                  & 603.bwaves\_s                          & 224               & 263.0    &                                    &                                                                               \\ \cline{3-5}
                                                  &                                                                                                                                  & 619.lbm\_s                             & 182               & 28.8   &                                    &                                                                               \\ \cline{3-5}
                                                  &                                                                                                                                  & 621.wrf\_s                             & 522               & 25.4   &                                    &                                                                               \\ \cline{3-5}
                                                  &                                                                                                                                  & 627.cam4\_s                            & 155               & 57.2   &                                    &                                                                               \\ \cline{3-5}
                                                  &                                                                                                                                  & 628.pop2\_s                            & 532               & 22.3   &                                    &                                                                               \\ \cline{3-5}
                                                  &                                                                                                                                  & 638.imagick\_s                         & 507               & 28.4   &                                    &                                                                               \\ \cline{3-5}
                                                  &                                                                                                                                  & 644.nab\_s                             & 191               & 91.5   &                                    &                                                                               \\ \cline{3-5}
                                                  &                                                                                                                                  & 649.fotonik3d\_s                       & 244               & 37.3   &                                    &                                                                               \\ \cline{3-5}
                                                  &                                                                                                                                  & 654.roms\_s                            & 245               & 64.2   &                                    &                                                                               \\ \cline{3-6}
                                                  &                                                                                                                                  & 600.perlbench\_s                       & 832               & 2.1    &  \multicolumn{1}{c|}{\multirow{-10}{*}{48.7 (FP)}}                         &                                                                               \\ \cline{3-5}
                                                  &                                                                                                                                  & 602.gcc\_s                             & 823               & 4.8    &                                   &                                                                               \\ \cline{3-5}
                                                  &                                                                                                                                  & 605.mcf\_s                             & 1369              & 3.5    &                                   &                                                                               \\ \cline{3-5}
                                                  &                                                                                                                                  & 620.omnetpp\_s                         & 815               & 2.0    &                                      &                                                                               \\ \cline{3-5}
                                                  &                                                                                                                                  & 623.xalancbmk\_s                       & 444               & 3.2    &                                   &                                                                               \\ \cline{3-5}
                                                  &                                                                                                                                  & 625.x264\_s                            & 703               & 2.5    &                                   &                                                                               \\ \cline{3-5}
                                                  &                                                                                                                                  & 631.deepsjeng\_s                       & 651               & 2.2    &                                    &                                                                               \\ \cline{3-5}
                                                  &                                                                                                                                  & 641.leela\_s                           & 999               & 1.7    &                                   &                                                                               \\ \cline{3-5}
                                                  &                                                                                                                                  & 648.exchange2\_s                       & 807               & 3.6    &                                   &                                                                               \\ \cline{3-5}
\multirow{-19}{*}{SPECspeed}                      & \multirow{-19}{*}{\begin{tabular}[c]{@{}l@{}}Unix\\ (AIX,\\  HP-UX,\\  Linux,\\  Mac OS X,\\  Solaris),\\  Windows~\cite{SPECCPU2017}\end{tabular}} & 657.xz\_s                              & 492               & 12.6    &  \multicolumn{1}{c|}{\multirow{-10}{*}{3.2 (INT)}}                                 & \multirow{-19}{*}{\begin{tabular}[c]{@{}l@{}}Higher scores mean that \\ less time is needed~\cite{SPECCPU2017}\end{tabular}} \\ \hline
                                                  &                                                                                                                                  & blackscholes                           & 133           & \multicolumn{1}{c|}{}                      & 133 & \multicolumn{1}{c|}{}                                                         \\ \cline{3-4} \cline{6-6}
                                                  &                                                                                                                                  & bodytrack                              & 346           & \multicolumn{1}{c|}{}               & 346      & \multicolumn{1}{c|}{}                                                         \\ \cline{3-4} \cline{6-6}
                                                  &                                                                                                                                  & canneal                                & 258           & \multicolumn{1}{c|}{}          & 258            & \multicolumn{1}{c|}{}                                                         \\ \cline{3-4} \cline{6-6}
                                                  &                                                                                                                                  & facesim                                & 771           & \multicolumn{1}{c|}{}            & 771          & \multicolumn{1}{c|}{}                                                         \\ \cline{3-4} \cline{6-6}
                                                  &                                                                                                                                  & fluidanimate                           & 974           & \multicolumn{1}{c|}{}              & 974        & \multicolumn{1}{c|}{}                                                         \\ \cline{3-4} \cline{6-6}
                                                  &                                                                                                                                  & freqmine                               & 776           & \multicolumn{1}{c|}{}             & 776        & \multicolumn{1}{c|}{}                                                         \\ \cline{3-4} \cline{6-6}
                                                  &                                                                                                                                  & streamcluster                          & 2037          & \multicolumn{1}{c|}{}               & 2037       & \multicolumn{1}{c|}{}                                                         \\ \cline{3-4} \cline{6-6}
                                                  &                                                                                                                                  & swaptions                              & 424          & \multicolumn{1}{c|}{}            & 424          & \multicolumn{1}{c|}{}                                                         \\ \cline{3-4} \cline{6-6}
                                                  &                                                                                                                                  & x264                                   & 144           & \multicolumn{1}{c|}{}            & 144          & \multicolumn{1}{c|}{}                                                         \\ \cline{3-4} \cline{6-6}
                                                  &                                                                                                                                  & {\color[HTML]{333333} dedup}           & 58            & \multicolumn{1}{c|}{}            & 58         & \multicolumn{1}{c|}{}                                                         \\ \cline{3-4} \cline{6-6}
                                                  &                                                                                                                                  & {\color[HTML]{333333} raytrace}        & 245          & \multicolumn{1}{c|}{}            & 245         & \multicolumn{1}{c|}{}                                                         \\ \cline{3-4} \cline{6-6}
\multirow{-12}{*}{PARSEC3.0}                      & \multirow{-12}{*}{\begin{tabular}[c]{@{}l@{}}Linux/i386, \\ Linux/AMD64,\\  Linux/Itanium,\\  Solaris/Sparc~\cite{bienia2008parsec}\end{tabular}}     & {\color[HTML]{333333} vips}            & 179           & \multicolumn{1}{c|}{\multirow{-12}{*}{\diagbox{}{}}} & 179 & \multicolumn{1}{c|}{\multirow{-12}{*}{\diagbox{}{}}}                                    \\ \hline
                                                  &                                                                                                                                  & 400.perlbench                          & 742               & 13.2 &                                       &                                                                               \\ \cline{3-5}
                                                  &                                                                                                                                  & 401.bzip2                              & 603               & 16.0 &                                        &                                                                               \\ \cline{3-5}
                                                  &                                                                                                                                  & 403.gcc                                & 373               & 21.6 &                                      &                                                                               \\ \cline{3-5}
                                                  &                                                                                                                                  & 429.mcf                                & 283               & 32.2 &                                        &                                                                               \\ \cline{3-5}
                                                  &                                                                                                                                  & 445.gobmk                              & 567               & 18.5 &                                      &                                                                               \\ \cline{3-5}
                                                  &                                                                                                                                  & 456.hmmer                              & 433               & 21.6 &                                      &                                                                               \\ \cline{3-5}
                                                  &                                                                                                                                  & 458.sjeng                              & 880               & 13.7 &                                      &                                                                               \\ \cline{3-5}
                                                  &                                                                                                                                  & 462.libquantum                         & 409               & 50.6 &                                      &                                                                               \\ \cline{3-5}
                                                  &                                                                                                                                  & 464.h264ref                            & 983               & 22.5  &                                     &                                                                               \\ \cline{3-5}
                                                  &                                                                                                                                  & 471.omnetpp                            & 434               & 14.4  &                                     &                                                                               \\ \cline{3-5}
\multirow{-11}{*}{\begin{tabular}[c]{@{}l@{}}CINT2006 \\ (speed)\end{tabular}}                    & \multirow{-11}{*}{\begin{tabular}[c]{@{}l@{}}Unix\\ (AIX,\\  HP-UX,\\  Linux,\\  Mac OS X,\\  Solaris),\\  Windows~\cite{CPU2006}\end{tabular}} & 473.astar                              & 553               & 12.7       &  \multicolumn{1}{c|}{\multirow{-11}{*}{19.6}}                              & \multirow{-11}{*}{\begin{tabular}[c]{@{}l@{}}Higher scores mean that \\ less time is needed~\cite{SPECCPU2017}\end{tabular}}                \\ \hline
                                                  &                                                                                                                                  & 410.bwaves                             & 406               & 33.5 &                                      &                                                                               \\ \cline{3-5}
                                                  &                                                                                                                                  & 433.milc                               & 549               & 16.7 &                                      &                                                                               \\ \cline{3-5}
                                                  &                                                                                                                                  & 434.zeusmp                             & 400               & 22.8 &                                      &                                                                               \\ \cline{3-5}
                                                  &                                                                                                                                  & 435.gromacs                            & 334               & 21.4 &                                      &                                                                               \\ \cline{3-5}
                                                  &                                                                                                                                  & 436.cactusADM                          & 385               & 31.1 &                                      &                                                                               \\ \cline{3-5}
                                                  &                                                                                                                                  & 437.leslie3d                           & 229               & 41.0  &                                     &                                                                               \\ \cline{3-5}
                                                  &                                                                                                                                  & 444.namd                               & 485               & 16.5 &                                      &                                                                               \\ \cline{3-5}
                                                  &                                                                                                                                  & 450.soplex                             & 278               & 30.0 &                                      &                                                                               \\ \cline{3-5}
                                                  &                                                                                                                                  & 453.povray                             & 276               & 19.2 &                                      &                                                                               \\ \cline{3-5}
                                                  &                                                                                                                                  & 454.calculix                           & 963               & 8.57  &                                     &                                                                               \\ \cline{3-5}
                                                  &                                                                                                                                  & 459.GemsFDTD                           & 354               & 30.0 &                                      &                                                                               \\ \cline{3-5}
                                                  &                                                                                                                                  & 465.tonto                              & 497               & 19.8 &                                      &                                                                               \\ \cline{3-5}
                                                  &                                                                                                                                  & 470.lbm                                & 337               & 40.8  &                                     &                                                                               \\ \cline{3-5}
\multirow{-14}{*}{\begin{tabular}[c]{@{}l@{}}CFP2006 \\ (speed)\end{tabular}}                     & \multirow{-14}{*}{\begin{tabular}[c]{@{}l@{}}Unix\\ (AIX,\\  HP-UX,\\  Linux,\\  Mac OS X,\\  Solaris),\\  Windows~\cite{CPU2006}\end{tabular}} & 482.sphinx3                            & 665               & 29.3       &\multicolumn{1}{c|}{\multirow{-14}{*}{23.9}}                                  & \multirow{-14}{*}{\begin{tabular}[c]{@{}l@{}}Higher scores mean that \\ less time is needed~\cite{SPECCPU2017}\end{tabular}}                \\ \hline
\end{tabular}
\end{table*}

Employing these benchmark suites, we evaluate the performance of a subject,  the Intel Xeon Gold 5120T processor,  and proceed to compare the resultant evaluation outcomes. During the experiments, apart from the processor itself, we provide the support system in the following manner: a 384 GB of memory, a 16TB disk, and the utilization of Ubuntu 20.04 as the operating system. To facilitate the compilation process, we employ the GNU Compiler Collection (GCC) version 9.4. We also use the largest data set of each benchmark suite (for SPEC CPU, it is a `ref' data set, and for PARSEC, it is a `native' data set) and run each workload three times for a comprehensive evaluation. For the SPEC CPU2006 benchmark suite, we use the default configuration file of SPECspeed Metric.

The evaluation outcomes are presented in Table~\ref{CPU_Benchmark_Evaluation}. The discernible discrepancies observed in the evaluation outcomes can be comprehensively elucidated by taking into account the significant disparities inherent in different benchmark suites. The variations encompass the selection of distinct problem or task instances, the algorithms, the implementation of algorithms,  the value functions utilized, the composite metrics employed for evaluation, the reference support system, and the reference subject, which is witnessed by  Table~\ref{CPU_Evaluation_Conditions}.




The experimental findings illustrated in Table \ref{CPU_Benchmark_Evaluation} reveal significant discrepancies in the evaluation outcomes of the same CPU when assessed using different CPU benchmark suites (the SPEC CPU and PARSEC benchmark suites). Furthermore, comparing evaluation outcomes from different evaluators becomes challenging when they employ different benchmarks. 
There are several reasons as follows. Firstly, they utilize different value functions. Secondly, the SPEC CPU benchmark suite encompasses a reference machine, whereas the PARSEC benchmark suite lacks such a basis for comparison. Moreover, the EECs cannot be ensured even when using the same value function and reference machine. This is because different benchmark suites introduce variations in problem or task instances, algorithms, and the implementation of algorithms.

Moreover, there are significant differences in evaluation outcomes of the same CPU, even when using the same benchmark suite with different versions. 
It stems from the different implementations of algorithms on varied support systems. Let us take the gcc workload in the SPEC CPU benchmark suite as an example. When evaluated under the CPU2006 benchmark suite, the CPU achieved a score of 21.6 in the 403.gcc workload, while it scored 4.8 in the 602.gcc\_s workload under the SPECspeed benchmark of SPEC CPU2017. These scores show a disparity of nearly five-fold. The discrepancies in evaluation outcomes can be mainly attributed to variations in the reference machine used by the respective benchmark suites, as outlined comprehensively in Table ~\ref{CPU_Evaluation_Conditions}. Moreover, the two workloads utilize different GCC compiler versions, with the 403.gcc workload utilizing GCC version 3.2 and the 602.gcc\_s workload utilizing GCC version 4.5. Although the command flag of both workloads is `ref,' the input data of the 403.gcc workload consists of nine C-code workloads, while the input data of the 602.gcc\_s workload is the preprocessed GCC compiler code. Additionally, 403.gcc is not multi-threaded, while the multi-threaded is permitted for 602.gcc\_s.

Furthermore, when being implemented with the same version of GCC compiler and using the same input data, the variances in evaluation outcomes for the gcc workload between the 502.gcc\_r workload in SPECrate benchmark suite and  602.gcc\_s in SPECspeed benchmark suite are more than twentyfold, which stem from the adoption of disparate value functions and the distinct implementations of the same algorithm. The SPECrate benchmark suite workloads are designed to assess throughput, employing multiple copies of a single-thread implementation during evaluations, while the SPECspeed benchmark suite workloads solely measure execution time, and the utilization of multiple threads is optional throughout the evaluation process. For the evaluation condition, 502.gcc\_r workload makes fifty-six copies, running with a single thread, while 602.gcc\_s workload has only one copy but runs with fifty-six threads.

The observed variations in evaluation outcomes for a particular CPU across different benchmark suites underscore the necessity of advancing the science and engineering of evaluation. While state-of-the-art CPU benchmarks have made significant progress, they do have certain drawbacks that need to be addressed. Firstly, the lack of comparability among evaluation results from different evaluators is a significant concern. Secondly, the significant discrepancies in evaluation outcomes can not be traceable. Lastly, state-of-the-art CPU benchmarks often struggle to provide a realistic estimate of the parameters of real-world systems (ES) with a high level of confidence. 

According to the comprehensive elements of a benchmark discussed in Section~\ref{essence_benchmark}, many CPU benchmarks, such as the SPEC CPU benchmark suites, primarily focus on the reference implementation of algorithms, metrics, and references while neglecting other essential constituents and components. In the following section, we will carefully analyze and highlight the shortcomings of state-of-the-art and state-of-the-parctice evaluation and benchmarks, employing our own terminology.

\subsection{The reflections on state-of-the-art and state-of-the-practise benchmarks and evaluation }

\begin{figure*} 
	\centering
		\includegraphics[scale=.5]{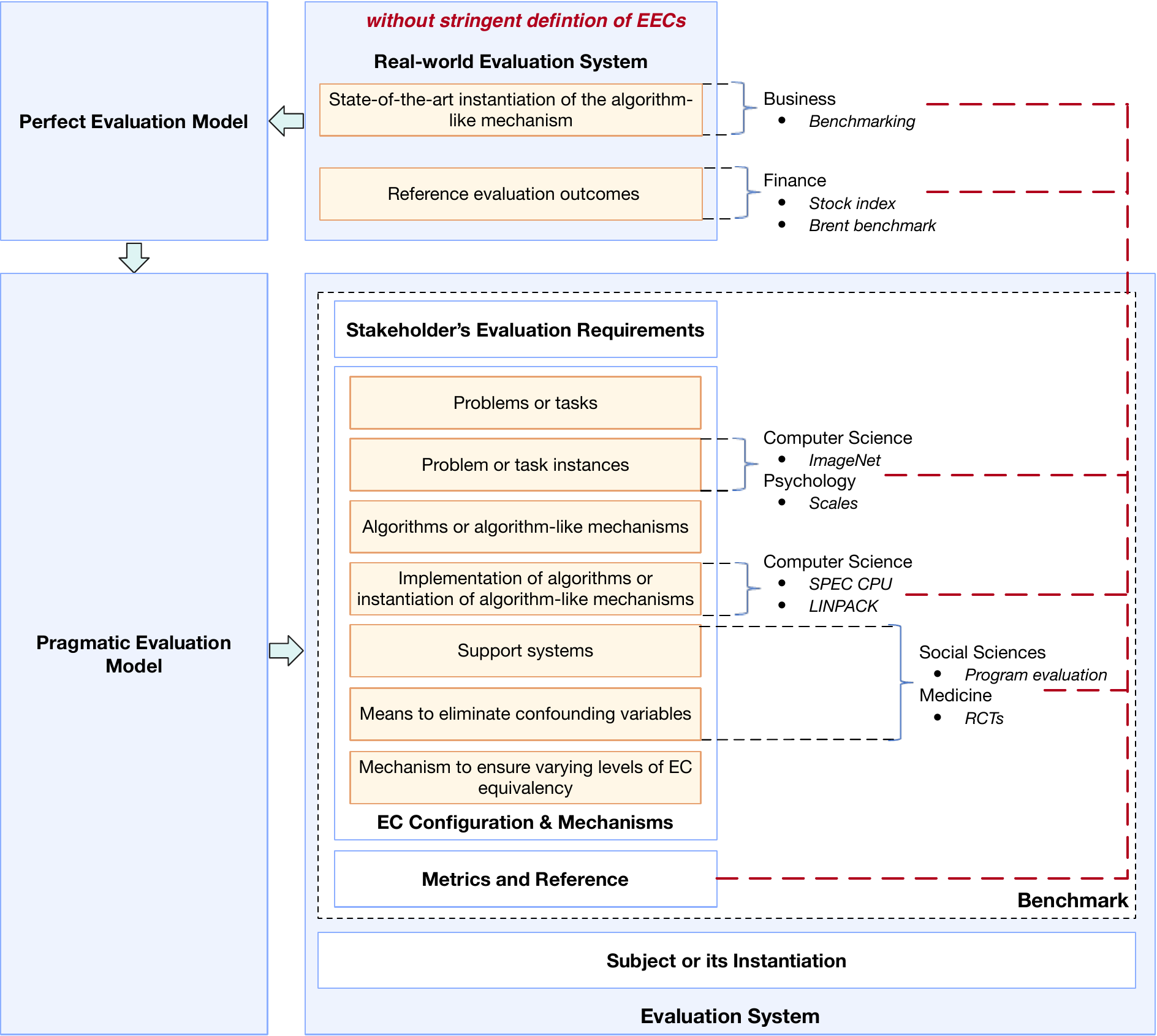}
	\caption{The reflections on state-of-the-art and state-of-the-practice benchmarks and evaluation are based on the science and engineering of evaluation.}
	\label{limitation_state_of_art_benchmark}
\end{figure*}

 
 To further illustrate the limitations of existing evaluation and benchmarking practices, we present Figure~\ref{limitation_state_of_art_benchmark}, which showcases these shortcomings within the evaluatology framework. By examining this figure, we can gain a clearer understanding of the areas where state-of-the-art and state-of-the-practice evaluation and benchmarks fall short.





It is evident that \textit{a lack of consensus exists regarding concepts and terminologies across different areas of study}. This lack of consensus often leads to confusion and misinterpretation, especially when the same terms are used in different disciplines with varying meanings.

For example, the term "benchmark" is commonly employed in computer science, finance, and business disciplines but without a formal definition. Moreover, even within these fields, the definition of "benchmark" can be vague and subject to interpretation. In contrast, psychology may use the term "scale" as a concept similar to benchmark, while social science and medicine may not have an analogous concept at all.

Recognizing this challenge, our work has aimed to propose universal concepts and terminologies that can bridge these disciplinary gaps. By establishing clear and standardized definitions, we seek to promote a shared understanding and facilitate effective communication and collaboration across different areas of study.

\textit{Few works discuss the essence of evaluation, let alone reaching a consensus on it.} Evaluation is often mistakenly equated with measurement or testing without clear differentiation. For instance, in computer science and psychology, evaluation and measurement are often used interchangeably. In the context of testing, where the goal is to determine whether an individual or a system aligns with the expected behavior defined by test oracles, evaluation is often conflated with testing. For instance, according to the SPEC terminology, a benchmark refers to "a test, or set of tests, designed to compare the performance of one computer system against the performance of others"~\cite{kounev2020systems, SPEC_glossary}. SPEC is a highly influential benchmark organization. Our work has revealed the essence of the evaluation. 

\textit{The evaluation theories and methodologies proposed are often domain-specific, with a lack of universally applicable foundational principles and evaluation methodologies that transcend diverse disciplines.} Different disciplines do not delve into the underlying principles of evaluation. Instead, they adopt a pragmatic approach and prioritize guidelines for conducting evaluations within specific contexts.

 For instance, in the medical discipline, the focus is primarily on eliminating confounding variables within the specific groups or cohorts being studied. In the business discipline, efforts are concentrated on searching the state of the practice.

The most rigorous theoretical foundation can be found in the field of clinical trials. For instance, Randomized Controlled Trial (RCT) techniques are employed to rule out the effect of confounding variables. However, there is a lack of universal problem formulations or fundamental solutions that fully consider the intricate interactions among the key components of EMs in diverse scenarios.

There are two serious drawbacks to the RCT methodology and its variants. Firstly, there is a lack of a stringent hierarchical definition of EC and EEC.  The variations in ECs can introduce confounding that may affect the results and make meaningful comparisons difficult. Without ensuring EECs, it becomes an illusion to expect comparable evaluation outcomes.

Secondly, when it comes to studying complex systems such as human beings or experimental animals, which we refer to as support systems, the RCT methodology and its variants may struggle to establish an REM. This kind of support system is characterized by a multitude of independent variables, making it difficult to isolate and control all relevant factors in a controlled experimental setting. Consequently, it becomes challenging to eliminate confounding variables and ensure unbiased evaluation outcomes completely.

In the realms of business and finance, different observational study methodologies are widely used, and we have revealed their inherent limitations in Section~\ref{basic_concept_definition}. An observational study is not even an experiment. Certainly, it can not eliminate confounding variables and reveal the cause-and-effect relationships.  In the business discipline, benchmarking assumes the state-of-the-art instantiation of the algorithm-like mechanism and the reference evaluation outcomes. 
In finance and education disciplines, benchmarks or indexes assume the role of reference evaluation outcomes in an observational study that measures variables of interest but does not attempt to influence the response~\cite{statics_book}. 

Rossi et al.~\cite{rossi2018evaluation} propose a valuable framework for evaluating methodologies in the field of social science. However, they do not provide a universal theory that can be applied to different disciplines. Their limitations stem from their narrow focus on assessing social programs without developing a generalized theory for evaluating other subjects in complex conditions.

Rossi et al. indeed utilized or developed some approaches to isolate the social programs' impacts, e.g., comparison group designs and randomized controlled trials (RCT), but they failed to explicitly state the underlying principles and methodology for universal science and engineering of evaluation. 

Within the computer science field, there are varying viewpoints and perspectives. For example, Hennessy et al.\cite{hennessy2011computer} highlight the significance of benchmarks and define them as programs specifically selected for measuring computer performance. On the other hand, John et al.\cite{john2018performance} compile a book on performance evaluation and benchmarking without providing formal definitions for these concepts. Kounev et al.\cite{kounev2020systems} present a formal definition of benchmarks as "tools coupled with methodologies for evaluating and comparing systems or components based on specific characteristics such as performance, reliability, or security." The ACM SIGMETRICS group\cite{browne1975analysis,knudson1985performance} considers performance evaluation as the generation of data that displays the frequency and execution times of computer system components, with a preceding orderly and well-defined set of analysis and definition steps.

In psychology, social and personality psychologists often utilize scales, such as psychological inventories, tests, or questionnaires, to assess psychometric variables~\cite{furr2011scale, furr2011scale}.  While these tools are commonly used, it is important to recognize that they rely on virtual assessments and self-report-style evaluations, which may introduce potential distortions. 

To overcome this limitation, we suggest implementing a physical application of an EC to the subjects, supplemented with a variety of measurement instruments. This approach aims to provide a more objective and accurate assessment of various aspects, including attitudes, traits, self-concept, self-evaluation, beliefs, abilities, motivations, goals, and social perceptions~\cite{furr2011scale}, by incorporating tangible and observable data.

\textit{Various disciplines have proposed engineering approaches to evaluations. However, they fail to provide universal benchmark concepts, theories, principles, and methodologies.} 

For instance, benchmarks are commonly utilized in finance, computer science, and business, albeit with inconsistent meanings and practices. 
Regrettably, there have been limited discussions in previous works regarding universal benchmark principles and methodologies that can be applied across different disciplines.
From a computer science standpoint, Kounev et al.~\cite{kounev2020systems} provide a comprehensive foundation for benchmarking, including metrics, statistical techniques, experimental design, and more. 

Most state-of-the-art and state-of-the-practice benchmarks overlook an essential aspect: the stakeholders' evaluation requirements. This oversight leads to a failure to consider different and diverse evaluation requirements. For instance, they do not enforce the discrepancy threshold in evaluation outcomes, nor do they consider evaluation confidence, among other crucial factors. As a result, most CPU benchmarks are ill-equipped to meet the evaluation requirements in scenarios involving safety-critical, mission-critical, and business-critical applications.

Another issue is the lack of a stringent definition for similar concepts, such as an EEC or LEEC. For example, most CPU or AI (deep learning) benchmarks, like ImageNet, fail to provide a clear definition of an EEC or LEEC. Instead, they jump directly into the implementation of algorithms or a specific dataset labeled with the ground truth without proper justification. Additionally, the support system, which plays a crucial role in some cases, is omitted without any explanation of the condition of simplifying the benchmarks. Furthermore, most of the methodologies fail to discuss the confounding elimination mechanism. This oversight can potentially introduce bias and inaccuracies in the evaluation outcomes.



Not surprisingly, the intricate evaluation mechanisms and policies introduced in Section~\ref{Phase_Two} are not explicitly discussed in the design and implementation of most benchmarks. For instance, it fails to address important aspects such as investigating and characterizing real-world ES, the design and implementation of a perfect EM, the modeling policy and procedure from a real-world ES to an EM, and the sampling policy and procedure from a perfect EC to a pragmatic EC. This omission makes it difficult for the benchmark to adapt to intricate evaluation scenarios. 

It is crucial to include these mechanisms and policies to ensure the benchmark's applicability and effectiveness in complex evaluation scenarios.
Without explicit discussion of the real-world ES, it is difficult to establish an EC that captures the characteristics and requirements of real-world evaluations.
Furthermore, exploring different sampling and modeling policies is essential to gain the confidence of the evaluation community in using the benchmark for inferring parameters of real-world ES. By carefully designing these policies, we can strike a balance between achieving high accuracy in evaluation outcomes and managing the associated evaluation costs.

There are many widely used AI (deep learning) benchmarks. 
Taking the ImageNet dataset as an illustrative example~\cite{deng2009imagenet}, we reveal their limitations. Firstly, a specific AI benchmark like ImageNet can not be traced back to an explicit formulation of a problem or task and instead manifests itself in the form of a dataset containing ground truth, which may possess certain biases. In other scenarios, we also encounter challenges in identifying a precise mathematical function that accurately models the chemical and biological activities within the human body (Case Three in Section~\ref{revisiting}) or the social dynamics within the target population (Case Four in Section~\ref{revisiting}). Secondly, the benchmark relies on an unverified assumption that the data distribution within the real world closely aligns with that of the collected dataset to a considerable extent. Thirdly, in real-world applications, we use the statistic of a sample---a specific benchmark--- to infer the parameters of the entire population. However, we do not know their confidence levels and intervals.

\subsection{What is the benefit of evaluatology?}~\label{benefit_evaluationology}

Evaluatology serves as the foundational theory that encompasses evaluations in various fields of study. It provides a universal framework for optimizing the evaluation process, with four fundamental axioms serving as its basis. Formulating the core evaluation issues mathematically presents opportunities for seeking optimal solutions based on theoretical grounds. As a subdivision of evaluatology, benchmarkology offers a comprehensive engineering approach and methodology for evaluation, applicable across diverse disciplines.


Together, evaluatology and benchmarkology contribute reusable knowledge to different domains, encompassing universal terminology, principles, and methodologies. By sharing this knowledge base, they facilitate advancements in both the state-of-the-art and state-of-the-practice of evaluation in various realms. This unification of communities embarks on a collective journey to address future challenges.


\subsection{The differences between evaluation, measurement and testing}~\label{difference_between_measurement_evaluation} 

Drawing on the preceding analysis, this subsection elucidates the marked disparity between evaluation, measurement, and testing.


First and foremost, it is important to acknowledge that measurement or testing serves as a preliminary constituent within the broader framework of evaluation. In addition to measurement and testing, an evaluation encompasses a series of steps, which we have discussed in Section~\ref{evaluationonology}.

Furthermore, it is crucial to recognize that the measurement results are of an objective nature, assuming the existence of an inherent truth value for each measured quantity. Similarly, testing results also possess an objective nature as they typically yield either a positive or negative outcome for each test conducted.  

Conversely, evaluation results possess a certain degree of subjectivity, such as the formulation of value functions based on the underlying measurement data, which we have discussed in the first evaluation axiom in Section~\ref{evaluation_axiom}.

By virtue of the aforementioned reasons, we can assert that metrology or testing serves as but one foundational aspect in the realm of evaluations.

\section{Conclusion}~\label{conclusion}  

This article formally introduces evaluatology, a discipline encompassing both the science and engineering of Evaluations. Our contributions are three-fold.

First, in order to promote consistency and facilitate cross-disciplinary understanding, we propose the adoption of universal evaluation concepts and terminologies centered around evaluation conditions. 


Secondly, we reveal the essence of evaluation and propose five evaluation axioms as the foundational evaluation theory. Furthermore, we introduce the universal evaluation theory, principles, and methodology that govern the field of evaluation. 

We create evaluation conditions with different levels of equivalency and apply them to diverse subjects to establish reference evaluation models that alter a single independent variable at a time while keeping all other variables as controls. We discover that the key to effective and efficient evaluations in various complex scenarios lies in establishing a series of evaluation models that maintain transitivity.

Third, building upon the science of engineering, We formally define a benchmark as a simplified and sampled evaluation condition that ensures different equivalency levels. We present a benchmark-based universal engineering of evaluation across different disciplines, which we refer to as benchmarkology.

 \section{Acknowledgments}

\printcredits




\bio{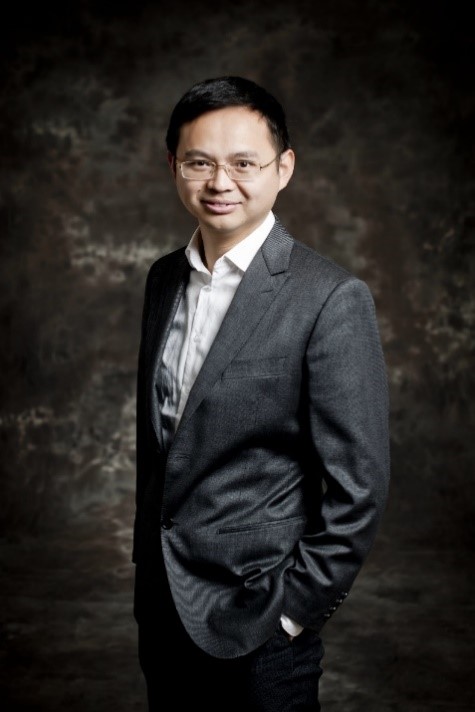}
Dr. Jianfeng Zhan is a Full Professor at the Institute of Computing Technology (ICT), Chinese Academy of Sciences (CAS), and University of Chinese Academy of Sciences (UCAS), the director of the Research Center for Advanced Computer Systems, ICT, CAS.  He received his B.E. in Civil Engineering and MSc in Solid Mechanics from Southwest Jiaotong University in 1996 and 1999 and his Ph.D. in Computer Science from the Institute of Software, CAS, and UCAS in 2002. His research areas span from Chips and systems to Benchmarks. A common thread is benchmarking, designing, implementing, and optimizing diverse systems. He has made substantial and effective efforts to transfer his academic research into advanced technology to impact general-purpose production systems. Several technical innovations and research results, including 35 patents from his team, have been adopted in benchmarks, operating systems, and cluster and cloud system software with direct contributions to advancing parallel and distributed systems in China or worldwide. Over the past two decades, he has supervised over ninety graduate students, post-doctors, and engineers. 
Dr. Jianfeng Zhan founded and chairs BenchCouncil and serves as the Co-EIC of TBench with Prof. Tony Hey. He has served as IEEE TPDS Associate Editor since 2018. He received the second-class Chinese National Technology Promotion Prize in 2006, the Distinguished Achievement Award of the Chinese Academy of Sciences in 2005, and the IISWC Best Paper Award in 2013, respectively. \endbio
\end{document}